\documentclass[conference]{IEEEtran}
\usepackage[T1]{fontenc}
\usepackage[scaled=0.8]{beramono} 
\usepackage[pdftex]{graphicx}
\usepackage{booktabs} 
\usepackage{multirow} 
\usepackage{xcolor} 
\usepackage{xspace} 
\usepackage[pagebackref=true]{hyperref} 
\usepackage{balance} 
\usepackage{wasysym} 
\usepackage{listings} 
\usepackage{ifthen}
\usepackage{mathptmx} 
\usepackage[tight,footnotesize]{subfigure}

\renewcommand*{\backref}[1]{}
\renewcommand*{\backrefalt}[4]{
   \ifcase #1
      No cited.
   \or
      (Cited on p.~#2)
   \else
      (Cited on pp.~#2)
   \fi}

\definecolor{darkblue}{rgb}{0,0,0.5}
\definecolor{lightgray}{rgb}{0.93,0.93,0.93}

\hypersetup{
	pdftitle={The Effect of DNS on Tor's Anonymity},
	pdfauthor={Benjamin Greschbach, Tobias Pulls, Laura M. Roberts, Philipp Winter, Nick Feamster},
	pdfkeywords={Tor, DNS, anonymity, traffic correlation, website fingerprinting},
	colorlinks=true,
	urlcolor=darkblue,
	linkcolor=darkblue,
	citecolor=darkblue
}

\newcommand{\ie}{{\it i.e.}\xspace}
\newcommand{\eg}{{\it e.g.}\xspace}
\newcommand{\ea}{{\it et al.}\xspace}
\newcommand{\etc}{{\it etc.}\xspace}
\newcommand{\first}{(\emph{i})\xspace}
\newcommand{\second}{(\emph{ii})\xspace}
\newcommand{\third}{(\emph{iii})\xspace}
\newcommand{\fourth}{(\emph{iv})\xspace}

\newcommand{\name}{DefecTor\xspace}

\newcommand{\TABLECAPTIONS}{top} 
\newcommand{\tabcaptext}{}
\ifthenelse{\equal{\TABLECAPTIONS}{top}}{%
\newcommand{\topcap}[1]{\caption{#1}}
\newcommand{\bottomcap}[1]{}
}
{
\newcommand{\topcap}[1]{}
\newcommand{\bottomcap}[1]{\caption{#1}}
}

\makeatletter
\def\blfootnote{\xdef\@thefnmark{}\@footnotetext}
\makeatother

\begin{document}

\title{The Effect of DNS on Tor's Anonymity}

\author{
\IEEEauthorblockN{Benjamin Greschbach$^*$}
\IEEEauthorblockA{\emph{KTH Royal Institute of Technology}}
\and
\IEEEauthorblockN{Tobias Pulls$^*$}
\IEEEauthorblockA{\emph{Karlstad University}}
\and
\IEEEauthorblockN{Laura M.  Roberts$^*$}
\IEEEauthorblockA{\emph{Princeton University}}
\and
\IEEEauthorblockN{Philipp Winter$^*$}
\IEEEauthorblockA{\emph{Princeton University}}
\and
\IEEEauthorblockN{Nick Feamster}
\IEEEauthorblockA{\emph{Princeton University}}
}


\maketitle

\begin{abstract}
Previous attacks that link the sender and receiver of traffic in the Tor network
(``correlation attacks'') have generally relied on analyzing traffic from TCP
connections. The TCP connections of a typical client application, however, are
often accompanied by DNS requests and responses. This additional traffic
presents more opportunities for correlation attacks.
This paper quantifies how DNS traffic can make Tor users more vulnerable to
correlation attacks.  We investigate how incorporating DNS traffic can make
existing correlation attacks more powerful and how DNS lookups can leak
information to third parties about anonymous communication. We \first develop a
method to identify the DNS resolvers of Tor exit relays; \second develop a new
set of correlation attacks (\name attacks) that incorporate DNS traffic to
improve precision; \third analyze the Internet-scale effects of these new
attacks on Tor users; and \fourth develop improved methods to evaluate
correlation attacks.
First, we find that there exist adversaries who can mount \name attacks: for
example, Google's DNS resolver observes almost 40\% of all DNS requests exiting
the Tor network. We also find that DNS requests often traverse ASes that the
corresponding TCP connections do not transit, enabling additional ASes to gain
information about Tor users' traffic.
We then show that an adversary who can mount a \name attack can often determine
the website that a Tor user is visiting with perfect precision, particularly for
less popular websites where the set of DNS names associated with that website
may be unique to the site.  We also use the Tor Path Simulator (TorPS) in
combination with traceroute data from vantage points co-located with Tor exit
relays to estimate the power of AS-level adversaries who might mount \name
attacks in practice. 
\end{abstract}

\blfootnote{$^*$All four authors contributed substantially, and share first authorship.}

\section{Introduction}
\label{sec:introduction}

We have yet to learn how to build anonymity networks that
resist global adversaries, provide low latency, and scale well.  Remailer
systems such as Mixmaster~\cite{mixmaster} and
Mixminion~\cite{Danezis2003a} eschew low latency in favor of
strong anonymity.
In contrast, Tor~\cite{dingledine2004a} trades off strong anonymity to
achieve low latency; Tor therefore
enables latency-sensitive applications such as web browsing but is
vulnerable to
adversaries that can observe traffic both
entering and exiting its network, thus enabling deanonymization.
Although Tor does not consider global adversaries in its threat model,
adversaries that can observe traffic for extended periods of time in
multiple network locations (\ie, ``semi-global'' adversaries) are a real
concern~\cite{Farrell2014a,Johnson2013a}; we need to better understand
the nature to which these adversaries exist in operational networks and
their ability to deanonymize users.

Past work has quantified the extent to which an adversary that
observes TCP flows between clients and servers (\eg, HTTP requests,
BitTorrent connections, and IRC sessions) can correlate traffic flows
between the client and the entry to the anonymity network and between
the exit of the anonymity network and its ultimate
destination~\cite{Johnson2013a,Murdoch2007a}. The ability to correlate
these two flows---a so-called {\em correlation attack}---can link the
sender and receiver of a traffic flow, thus compromising the anonymity
of both endpoints. Although TCP connections are an important part
of communications, the Domain Name System (DNS) traffic is also
quite revealing: for example, even loading a single webpage can generate
hundreds of DNS requests to many different domains. No previous analysis
of correlation attacks has studied how DNS traffic can exacerbate
these attacks.

\begin{figure}[t]
	\centering
	\includegraphics[width=0.65\linewidth]{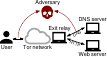}
	\caption{Past traffic correlation studies have focused on linking the TCP
		stream entering the Tor network to the one(s) exiting the network.  We
		show that an adversary can also link the associated DNS traffic, which
		can be exposed to many more ASes than the TCP stream.}
	\label{fig:overview}
\end{figure}

DNS traffic is highly relevant for correlation attacks because it
often traverses completely different paths
and autonomous systems (ASes) than the subsequent corresponding TCP
connections.  An attacker that can observe occasional DNS
requests may still be able to link both ends of the communication, even
if the attacker cannot observe TCP traffic between the exit of the
anonymity network and the server.
Figure~\ref{fig:overview} illustrates how an adversary may
monitor the connection between a user and the guard relay, and between the exit
relay and its DNS resolvers or servers.  This
territory---to-date, completely unexplored---is the focus of this work.

We first explore how Tor exit relays resolve DNS names.  By developing a new
method to identify all exit relays' DNS resolvers, we learn that Google
currently sees almost 40\% of all DNS requests exiting the Tor network.  Second,
we investigate which organizations can observe DNS requests that originate from
Tor exit relays.  To answer this question, we emulate DNS resolution for the
Alexa Top 1,000 domains from an autonomous system that is popular among exit
relays.  We find that DNS resolution for half of these domains traverses numerous
ASes that are not traversed for the subsequent HTTP connection to the web site.
We further introduce a new method to perform traceroutes from the networks where
exit relays are located, making our results significantly more accurate and
comprehensive than previous work.  Next, we show how the ability to observe DNS
traffic from Tor exit relays can augment existing website fingerprinting
attacks, yielding perfectly precise \name\footnote{The acronym is short for
\underline{D}NS-\underline{e}nhanced \underline{f}ingerprinting and
\underline{e}gress \underline{c}orrelation on \underline{Tor}.} attacks for
unpopular websites.  Finally, we use the Tor Path Simulator (TorPS)~\cite{TorPS}
to investigate the effects of Internet-scale \name attacks.

We demonstrate that DNS requests significantly increase the opportunity
for adversaries to perform correlation attacks. This finding should
encourage future work on correlation attacks to consider both TCP
traffic and the corresponding DNS traffic; future design decisions
should also be cognizant of this threat.  The measurement methods we use
to evaluate the effects of traffic correlation attacks are also more
accurate than past work. Our work \first serves as guidance to Tor exit
relay operators and Tor network developers, \second improves
state-of-the-art measurement techniques for analysis of correlation attacks, and \third
provides even stronger justification for introducing website fingerprinting defenses in
Tor.  To foster future work and facilitate the replication of
our results, we publish both our code and datasets.\footnote{Our project page is
available at \url{https://nymity.ch/tor-dns/}.}
In summary, we make the following contributions:
\begin{itemize}

\item We show how existing website fingerprinting attacks can be
  augmented with observed DNS requests by an AS-level adversary to
  yield perfectly precise \name attacks for unpopular websites.

\item We develop a method to identify the DNS resolver of exit
  relays. We find that Tor exit relays comprising 40\% of Tor's exit
  bandwidth rely on Google's public DNS servers to resolve DNS queries.

\item We quantify the extent to which DNS resolution exposes Tor users
  to additional AS-level adversaries who are not on the path between the
  sender and receiver.  We find that for the Alexa top 1,000 most
  popular websites, 60\% of the ASes that are on the paths between the
  exit relay and the DNS servers required to resolve the sites' domain
  names are not on the path between the exit relay and the website.

\item We develop a new measurement method to evaluate the extent to
  which ASes are on-path between exit relays and DNS resolvers. We use
  the RIPE Atlas~\cite{atlas} platform to achieve previously
  unprecedented path coverage and accuracy for evaluating the
  capabilities of AS-level adversaries.
\end{itemize}
\noindent
The rest of this paper is organized as follows.
Section~\ref{sec:background} presents background, and
Section~\ref{sec:related_work} relates our study to previous work.  In
Section~\ref{sec:landscape}, we shed light on the landscape of DNS in
Tor.  Section~\ref{sec:attack} discusses our \name attacks, which we
evaluate in Section~\ref{sec:analysis}.  We then model the
Internet-scale effect of our attacks in
Section~\ref{sec:internet-scale}.  Finally, we discuss our work in
Section~\ref{sec:discussion} and conclude the paper in
Section~\ref{sec:conclusion}.

\section{Background}
\label{sec:background}
We provide an introduction to the Tor network, website fingerprinting
attacks, as well as how the Tor network implements DNS resolution.

\paragraph{The Tor network}
The Tor network is an overlay network that anonymizes TCP streams such as web
traffic.  As of August 2016, it comprises approximately 7,000 relays and about two
million users.  The hourly published network consensus summarizes all relays
that are currently online.  Clients send data over the Tor network by randomly
selecting three relays---typically called the guard, middle, and exit
relay---that then form a virtual tunnel called a \emph{circuit}.  The guard
relay learns the client's IP address, but not its web activity, while the exit
relay gets to learn the client's web activity, but not its IP address.  Relays
and clients talk to each other using the Tor protocol, which uses 512-byte
\emph{cells}.  Finally, each relay is uniquely identified by its fingerprint---a
hash over its public key.

\paragraph{Website fingerprinting attacks}
The Tor network encrypts relayed traffic as it travels from the client
to the exit relay.  Therefore, intermediate parties such as the user's
Internet service provider (ISP) cannot read the contents of any packet.
Tor does not, however, protect other statistics about the network
traffic, such as packet inter-arrival timing, directions, and frequency.
The ISP can analyze these properties to infer the destinations that a
user is visiting.  The literature calls this attack \emph{website
fingerprinting}.

Past work evaluated website fingerprinting attacks in two settings:
a {\em closed-world} setting consists of a set of $n$ {monitored}
websites, and the attacker tries to learn which among all $n$ sites the
user is visiting with the notable restriction that the user can only
browse to one of the $n$ websites.  The {\em open-world} setting is more
realistic: the user can browse to {unmonitored} sites in addition to the
monitored sites. Unmonitored sites are, per definition, not known to the
attacker; thus, the attacker's traffic classifier cannot train on unmonitored
sites the user visits. The attacker's classifier can train on
whatever unmonitored sites it chooses, as long as the
classifier has not trained on an unmonitored site used for testing.
Two relevant metrics
in the open-world
setting are \emph{recall} and \emph{precision}.
Recall measures the probability that
a visit to a monitored site will be detected, while precision measures
the probability that a classification by the classifier of a visit to a
monitored site (positive test outcome) is the correct one. Consider a
classifier with 0.25 recall and 0.5 precision: on average, every fourth
visit by the user to a monitored site will be detected, and half of the
classifications by the classifier will be wrong. Errors can
either classify a monitored site as
unmonitored (lowering recall) or vice versa (lowering precision).
Mistaking one monitored site for another is less likely~\cite{Wang2015a}.

Wa-kNN is a website fingerprinting attack by Wang \ea~\cite{Wang2014a}
that uses a k-nearest neighbor classifier with a custom weight-learning
algorithm, WLLCC~\cite{Wang2015a}.  From {packet traces}
between a Tor client and its guard, Wa-kNN extracts a number of {features}
to classify each website.  Useful features include
the number of outgoing packets and bursts of packets in the same direction.
In the training phase, WLLCC adjusts {weights} of features extracted from
sites of known classes such that the {distance} between instances of the
same site (class) are minimized (collapsed).
In the testing phase, Wa-kNN determines the distance of a testing traffic trace
to all known training traces.  The distance calculation results in the $k$
nearest classes: if all classes are the same, then the testing trace is
classified as that class, otherwise it is classified as unmonitored.
In the open-world setting, one class represents all unmonitored sites both
during training and testing.  By increasing $k$, Wa-kNN can trade decreased
recall for increased precision.  We set $k=2$ when using Wa-kNN for higher
recall since \name is a highly precise attack.

Tor could eliminate website fingerprinting
attacks with encrypted, constant-bitrate channels between a Tor client and
its guard; other anonymity networks could use a similar technique.
Unfortunately, the Tor network's limited spare capacity does not
allow for such a throughput-intensive defense, but some research has
worked on making this type of defense
more efficient~\cite{Cai2014a,Juarez2016a,Wang2015a,adapativepadding}.

\paragraph{How Tor resolves DNS requests}
Tor clients must send DNS requests over Tor to prevent DNS {leakage}
(\ie, having a
DNS request travel over an unencrypted channel as opposed to over Tor
itself).  Tor does not transport UDP
traffic, but it implements a workaround to wrap DNS requests into Tor cells.
Using the
SOCKS protocol, applications can instruct the Tor client to establish a circuit to a
given domain and port.\footnote{SOCKS versions 4a and 5 support connection
initiation using domain names in addition to IP addresses.}
After the user types in a domain, say {\tt example.com}, the Tor browser establishes
a connection to the SOCKS proxy exposed by the local Tor client.
The Tor client then
selects an exit relay whose exit policy supports {\tt example.com} and port 443.
Next, the client sends a \texttt{RELAY\_BEGIN} Tor cell to the exit relay,
instructing it to first resolve {\tt example.com}, and then establish a TCP
connection to the resolved
address at port 443~\cite[\S~6.2]{tor-spec}.  After successfully establishing a
connection, the exit relay responds with a \texttt{RELAY\_CONNECTED}
cell.  The client can then exchange data with its intended
destination.  Another type of cell, \texttt{RELAY\_RESOLVE}, supports pure name
resolution, without establishing a subsequent TCP
connection~\cite[\S~6.4]{tor-spec}.  The exit relay responds with a
\texttt{RELAY\_RESOLVED} cell.

Exit relays send
their DNS requests to the system resolver, which Linux systems read from
\texttt{/etc/resolv.conf}.  Tor does not modify the system resolver and
uses whatever the exit relay operator configured, such as the ISP's resolver,
or public resolvers such as Google's {\tt 8.8.8.8} open public DNS resolver.
As of August 2016, exit relays cache DNS responses to speed up repeated lookups.
The caching layer for Tor clients, however, is off by default to prevent tracking attacks
due to modified DNS responses~\cite{nolocalcache}.

\section{Related Work}
\label{sec:related_work}

This paper combines traffic analysis methods for correlation attacks
with website fingerprinting attacks; we discuss related work in each of
these two areas.

\paragraph{Traffic analysis and correlation attacks}
Tor's threat model excludes global adversaries~\cite{dingledine2004a}, but the
practical threat of such adversaries is an important question that the
research community has
spent considerable effort on answering.  In 2004, when the Tor network comprised
only 33 relays, Feamster and Dingledine investigated the practical threat that
AS-level adversaries pose to anonymity networks~\cite{Feamster2004a}.
The authors considered an attacker that controls an AS
that is traversed both for ingress and egress traffic, allowing the
attacker to correlate both streams.  Using AS path prediction~\cite{Gao2001a},
Feamster and Dingledine found that powerful tier-1 ISPs reduce location
diversity of anonymity networks.  In 2007, Murdoch and Zieli\'{n}ski drew
attention to IXP-level adversaries, a class of adversaries that was missing in
Feamster and Dingledine's work~\cite{Murdoch2007a}.  Murdoch and Zieli\'{n}ski
showed that IXP adversaries are able to correlate traffic streams, even in the
presence of packet sampling rates as low as one in 2,000.

In 2013, Johnson
\ea~\cite{Johnson2013a} presented the first large-scale study on the risk of Tor
users facing relay-level and AS-level adversaries.  The authors developed
TorPS~\cite{TorPS} that simulates Tor circuits for a number
of user models.  By combining TorPS with AS path
prediction, Johnson \ea could answer questions such as the average time until a
Tor user's circuit is deanonymized by an AS or IXP.  Most recently in 2016,
Nithyanand \ea~\cite{Nithyanand2016a} used AS path prediction to evaluate the
practical threat faced by users in the top 10 countries using Tor.  In 2015,
Juen \ea~\cite{Juen2015a} examined the accuracy of path prediction algorithms
that prior work~\cite{Johnson2013a,Feamster2004a} used to estimate the threat of
correlation attacks.  The authors compared AS path predictions to millions of
traceroutes, initiated from 25\% of Tor relays by bandwidth at the
AS level, and found that
only 20\% of predicted paths matched the paths observed in traceroute.
Juen \ea could not consider the reverse path in traceroutes.  In 2015,
Sun \ea~\cite{Sun2015a} addressed this shortcoming; although past work treated
routing as static, Sun \ea showed that the dynamic nature of Internet
routing makes AS-level adversaries stronger than previous work had considered.

We improve on previous work in two significant ways: (\emph{i}) we are
the first to evaluate how DNS traffic exacerbates traffic correlation
attacks, both in concept and in practice; and (\emph{ii}) we develop and
deploy a scalable, sustainable version of the measurement method proposed by Juen
\ea~\cite{Juen2015a}.  Our method uses the volunteer-run RIPE Atlas
measurement platform~\cite{atlas}, as opposed to relying on relay operators
to run third-party scripts.  This approach allows us to fully automate
our method and achieve previously unprecedented scale.

\paragraph{Website fingerprinting}
In 2009, Herrmann \ea~\cite{Herrmann2009a} demonstrated the first website
fingerprinting attack against anonymity systems---including Tor---in a
closed-world setting.  In 2011, Panchenko \ea~\cite{Panchenko2011a} greatly
improved on Herrmann \ea's detection rate and provided insight into an open-world
setting.  In 2012, Cai \ea~\cite{Cai2012a} improved on previous work by
proposing an attack that used Hidden Markov Models to determine whether a sequence of
page requests all come from the same site.  The authors used an open-world
setting for their evaluation.  Wang and Goldberg~\cite{Wang2013a} proposed an
improved attack that employed a new method for data gathering.  In 2014, Wang
\ea~\cite{Wang2014a} further improved on their results with a
k-nearest neighbor classifier Wa-kNN and a custom weight-learning algorithm
(WLLCC~\cite{Wang2015a}) that in several rounds determine the optimal weights
for features extracted from traffic traces.
Cai \ea~\cite{Cai2014b}
determined which traffic features provide the most predictive power to detect
websites, proved a lower bound of any defense that achieves a certain level of
security, and provided a framework to investigate the performance of
website fingerprinting attacks.
Juarez~\cite{Juarez2014a} showed that all previous attacks
made several simplifying
assumptions; the work suggested that attacks are still
difficult to run outside a lab setting as an attacker will have to consider
operating system differences, page changes, and background traffic.
Recently, in 2016, Wang and Goldberg addressed many practical
roadblocks to website fingerprinting, such as noisy data and maintaining a training set,
further highlighting the need for website fingerprinting defenses in
Tor~\cite{Wang2016a}.

Panchenko \ea~\cite{Panchenko2016a} showed that web\emph{page}
fingerprinting (\ie, fingerprinting of any page on a site) is
significantly harder than web\emph{site} fingerprinting (\ie,
fingerprinting of only the start page of a site).  Hayes and Danezis
proposed k-fingerprinting, an attack with notably better performance
than Wa-kNN even in the face of defenses
\cite{Hayes2016a}. Their attack retains 30\% accuracy in a
closed-world setting against the WTF-PAD defense by Juarez et
al.~\cite{Juarez2016a}---a prime candidate for
deployment in Tor~\cite{adapativepadding}---at the cost of 50\% bandwidth
overhead. Juarez \ea used Wa-kNN to evaluate WTF-PAD and set $k=5$, as
recommended by Wang \ea for an optimal trade-off between recall and the
false positive rate.

In this paper, we show how to correlate and use observed DNS requests in
concert with website fingerprinting attacks,
which significantly
improves precision for web\emph{site} fingerprinting.
The two \name attacks that we present have implications
for the design of website fingerprinting defenses:
to mitigate \name attacks, open-world evaluations of the website fingerprinting
defense should minimize recall even when the website fingerprinting attack is
tuned to sacrifice precision for recall.  In the case of Wa-kNN, this means
a low $k$: our results are based on $k=2$.

\section{Understanding the Landscape}
\label{sec:landscape}

Before explaining our attack, we need to better understand how Tor performs DNS
resolution.  We begin by investigating how common it is for adversaries
to be able to observe DNS requests but \emph{not} subsequent TCP connections of
Tor users (Section~\ref{sec:as-exposure}).  We then seek to understand how
these results connect to the Tor network by determining the DNS resolvers used
by exit relays (Section~\ref{sec:mapping-resolvers}).

\subsection{Quantifying the additional AS exposure of DNS queries}
\label{sec:as-exposure}

Adversaries that can observe both DNS and subsequent TCP traffic (\eg, the ISP
of an exit relay) gain no benefit from seeing the client's DNS traffic,
since TCP traffic is sufficient to mount correlation
attacks~\cite{Murdoch2007a}.  In this work, we consider
adversaries that can observe traffic entering the Tor network and \emph{some}
DNS requests exiting the network---such as requests addressed to DNS
root servers---but {\em not} subsequent TCP traffic from exit relays.
We first determine the prevalence of these adversaries by measuring the
number of ASes that DNS queries traverse versus the
number of ASes subsequent web traffic traverses.

We quantify the {\em exposure} of DNS traffic versus TCP traffic as follows.  We
begin with Alexa's Top 1,000~\cite{alexatop1k}, a list of the 1,000 most popular
web sites as estimated by Alexa.  For each site, we conducted two
experiments.  First, we ran a TCP traceroute to the site, targeting port 80 to
mimic web traffic.  Second, we determined the DNS delegation path for the
website's DNS name using the {\tt dig} command's \texttt{+trace} feature.  The
delegation path of a domain name, say {\tt www.example.com}, is a list of
authoritative DNS servers, such as the authoritative server for {\tt .com}
pointing to the authoritative server of {\tt example.com}, which in turn points
to the authoritative server responsible for {\tt www.example.com}.  We also ran
UDP traceroutes to each server in the delegation path, targeting port 53 to
mimic DNS resolution.\footnote{The tool we developed for this purpose is
available online at \url{https://github.com/NullHypothesis/ddptr}.}
For both experiments, we then mapped all IP addresses in the traceroutes to AS
numbers~\cite{ipasn}, generating both a set of traversed ASes for DNS traceroutes
($\mathcal{D}$) and a set of traversed ASes for web traceroutes
($\mathcal{W}$).  Given these two sets for each of Alexa's Top
1,000, we compute the fraction of ASes that are \emph{only}
traversed for DNS traffic, but \emph{not} for web traffic ($\lambda$):

\begin{equation}
\label{equ:exposure}
\lambda \in [0, 1] =
\frac{|\mathcal{D} \setminus \mathcal{W}|}
     {|\mathcal{D} \cup \mathcal{W}|}.
\end{equation}
\noindent
The metric approaches 1 as the number of ASes that are only traversed for DNS
increases.  For example, if $\mathcal{D} = \{1,2,3\}$ and $\mathcal{W} =
\{2,3,4\}$, then $\lambda = \frac{|\{1\}|}{|\{1,2,3,4\}|} = \frac{1}{4} =
0.25$.  We determined $\lambda$ for each site in the Alexa Top 1,000.  We ran
the experiment on a French VPS in AS 16276, owned by OVH SAS.  We chose this AS
because, as of August 2016, it is the most popular AS by exit bandwidth, seeing
$10.98\%$ of exit traffic, closely followed by AS 12876 (owned by the French
Online S.A.S.) that sees $9.33\%$ of exit traffic.  The experiment resulted in
1,000~$\lambda$ values.  We repeated the experiment on a vantage point at our
institution, achieving similar results: while a two-sample Kolmogorov-Smirnov
test determined a statistically significant difference between both
distributions ($p \ll 0.01$), the medians (0.590 and 0.584) and standard
deviations (0.149 and 0.118) are similar.  Nevertheless, we believe that relays
in regions other than Western Europe or North America are likely to witness
significantly different exposure of DNS queries because many websites
outsource their DNS setup to providers such as CloudFlare whose points of
presence are centered around Western Europe and North America.

%
%

\begin{figure}[t]
	\centering
	\includegraphics[width=0.67\linewidth]{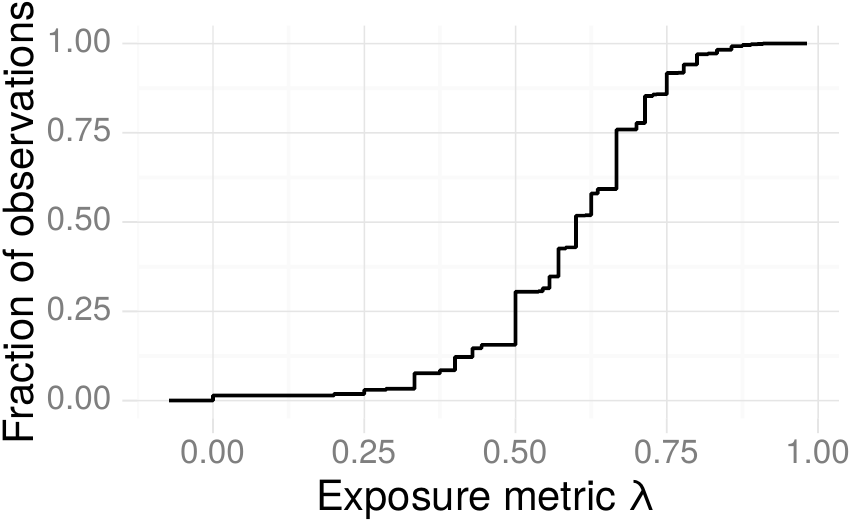}
	\caption{Our AS exposure metric $\lambda$ for Alexa's Top 1,000 web sites.
	For half of all sites, 60\% of ASes or more are only traversed for DNS
	resolution, but not for subsequent web traffic.}
	\label{fig:exposure}
\end{figure}

Figure~\ref{fig:exposure} shows the empirical CDF of all 1,000~$\lambda$ values
that we calculated for Alexa's Top 1,000 sites.  In total, this experiment
traversed 350 unique ASes for DNS requests and 339 unique ASes for web requests.
The figure shows that for half of Alexa's Top 1,000 domains, DNS-only ASes
account for 60\% or more of all traversed ASes.  This result
only applies to exit relays that do their own DNS resolution; for relays that
use a third-party resolver, the ASes that are traversed between
the exit relay and its DNS resolver is the metric of interest.  We conclude that adversaries that are
unable to observe a Tor user's TCP connection still have many opportunities to
see a TCP connection's corresponding DNS request.  Such adversaries include (\emph{i}) popular open
DNS resolvers such as Google and OpenDNS, (\emph{ii}) DNS root servers, and
(\emph{iii}) network adversaries located on the path to the previous two
entities.


\subsection{Determining how Tor exit relays resolve DNS queries}
\label{sec:mapping-resolvers}

Having shown that the Internet provides ample opportunity for
AS-level adversaries to snoop on DNS traffic from exit relays, we
now investigate how the exit relays in the Tor network resolve DNS
queries in practice. Before this study,
we only had anecdotal evidence (\eg, from OpenDNS-powered error
messages~\cite[\S~4.1]{Winter2014b}) that some exit relays would occasionally
show.

\begin{figure}[t]
	\centering
	\includegraphics[width=0.6\linewidth]{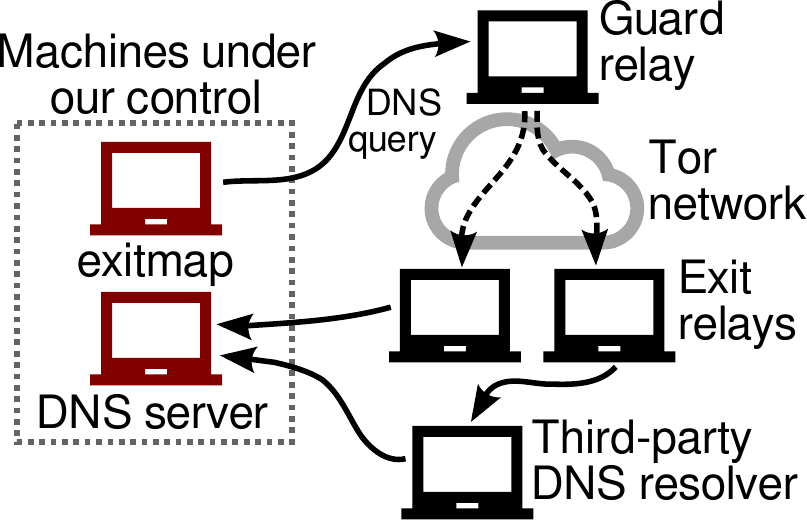}
	\caption{Our method to identify the DNS resolvers of exit relays.  Over
	each exit relay, we resolve relay-specific domain names that are under our
	control.  Inspecting our DNS server logs, we can then identify the IP
	address of all exit relay resolvers.}
	\label{fig:dnsenum}
\end{figure}

We identify the DNS resolver of all exit relays by using {\tt
exitmap}~\cite{exitmap}, a scanner for Tor exit relays.  {\tt Exitmap} automates
running a task such as fetching a webpage over all one thousand exit relays,
making it possible to see the Internet through the ``eyes'' of every single exit
relay.  Using {\tt exitmap}, we resolve unique, relay-specific domains over each
exit relay, to a DNS server under our control.  Figure~\ref{fig:dnsenum}
illustrates this experiment.  To improve reliability, we configured {\tt
exitmap} to use two-hop circuits instead of the standard three-hop circuits.
The first hop was a guard relay under our control.  Over each exit relay, we
resolved a unique domain {\tt PREFIX.tor.nymity.ch}.  The prefix consisted of
the relay's unique 160-bit fingerprint, concatenated to a random 40-bit string
whose purpose is to prevent caching, so exit relays indeed resolve each query
instead of responding with a cached element.  We controlled the authoritative
DNS server of {\tt tor.nymity.ch}, so we could capture both the IP address and
packet content of every single query for {\tt tor.nymity.ch}.

\begin{figure*}[t]
	\centering
	\includegraphics[width=\linewidth]{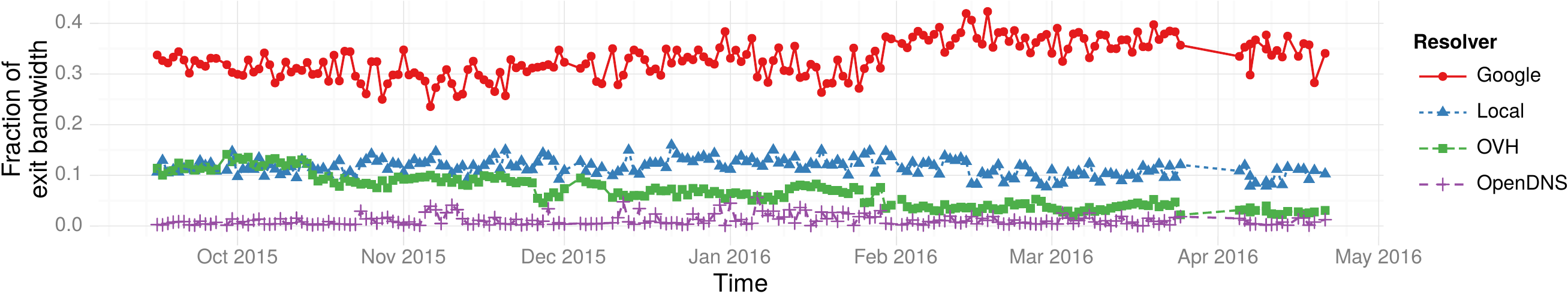}
	\caption{The popularity of some of the most popular DNS resolvers of exit
		relays over time.  The $y$ axis depicts the fraction of exit bandwidth
		that the respective resolver is responsible for.  Google's DNS resolver
		is by far the most popular, at times serving more than 40\% of all DNS
		requests coming out of the Tor network.  Google is followed by local
		resolvers, which average at around 12\%.  Once serving a fair amount of
		traffic, OVH dropped in popularity, and is now close to OpenDNS, an
		organization that runs an open resolver.}
	\label{fig:exit-resolvers}
\end{figure*}

An exit relay can either run its own resolver, as shown in the left exit relay in
Figure~\ref{fig:dnsenum}; or rely on a third-party resolver, such as the one
provided by its ISP, as shown in the right exit relay in Figure~\ref{fig:dnsenum}.  If
an exit relay runs its own resolver, we expect to receive a DNS request from
the exit relay's IP address, but if an exit relay uses a third-party resolver,
we expect to receive a request from an unrelated IP address.  Having encoded
relay-specific fingerprints in the query names, we are able to map queries to
exit relays in such cases.  We ran this experiment from September 2015 to May
2016, at least once a day.  Once we started to analyze our data, we identified
the following two challenges:

\begin{itemize}
\item {\em DNS proxies:}
We found that several exit relays used DNS
proxies---machines that passed on DNS requests to an actual resolver
instead of resolving it themselves.  Google's server {\tt 8.8.8.8} is a popular
example of a DNS proxy;
several dozen machines perform resolution in the
background~\cite{google-proxies}.  DNS proxies do
not interfere with our measurements, so we ignored them.

\item {\em Multiple DNS resolvers:}
On Linux systems, DNS resolution is controlled by the file
\texttt{/etc/resolv.conf}.  It contains a list of up to three DNS resolvers
that are queried in order.  If the primary resolver does not respond within a
time limit, the system falls back to the second, and finally the third
resolver.  Our data suggests that several exit relays used
different resolvers in subsequent {\tt exitmap} scans---one relay, for example,
used both Google's DNS resolver and one provided by its ISP.  For our
visualization, we only consider the first resolver we observed for an exit
relay, which might not be the primary resolver.
\end{itemize}
\noindent
Figure~\ref{fig:exit-resolvers} illustrates the fraction of DNS requests that
four of the most popular organizations could observe.  Google averages at 33\%,
but at times saw more than 40\% of all DNS requests exiting the Tor network---an
alarming number for a single organization.  Second to Google is ``Local''---exit
relays that run their own resolver, averaging at 12\%.  Next is OVH, which used
to be as popular as local resolvers, but slowly lost its share over time.  Note
that in contrast to Google, OVH does not run a public DNS server; the company's
resolvers are only accessible to its customers.  Finally, there is OpenDNS,
which also runs public DNS resolvers.  OpenDNS saw occasional spikes in
popularity but always remained in the single digits.  Apart from the illustrated
top resolver setups, the distribution has a long tail, presumably consisting of
many ISP resolvers.

\section{\name Attacks}
\label{sec:attack}

As with conventional correlation attacks, an attacker must observe
traffic that is both entering and exiting
the Tor network; in contrast to threat models from previous work, we
incorporate DNS instead of only
TCP traffic.
Figure~\ref{fig:attack-scenario} illustrates our correlation attack; it requires the
following building blocks:
\begin{figure}[t]
	\centering
	\includegraphics[width=0.8\linewidth]{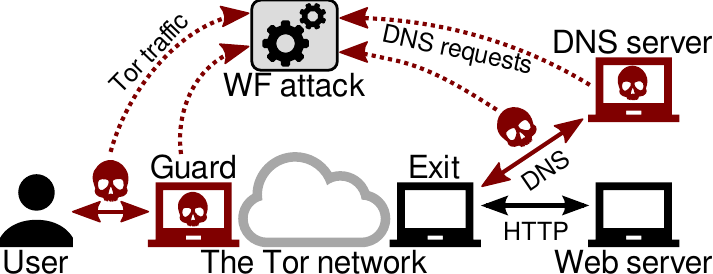}
	\caption{An overview of the \name attack.  An adversary must monitor
		both ingress (encrypted Tor traffic) and egress (DNS request) traffic.
		A AS-level adversary between the
		client and its guard monitors ingress traffic.  The same adversary
		monitors egress traffic between the exit and a DNS server, or the DNS
		server itself.  Both ingress and egress traffic then serve as input to the
		\name attack.}
	\label{fig:attack-scenario}
\end{figure}

\begin{itemize}
    \item \emph{Ingress sniffing:} An attacker must observe traffic that is
		entering the Tor network.  The attacker can operate on the network level,
		as a malicious ISP or an intelligence agency.  In addition, the
		attacker can operate on the relay level by running a malicious Tor guard
		relay.  In both cases, the attacker can only observe encrypted
		data, so packet lengths and
		directions are the main inputs for website fingerprinting~\cite{Panchenko2016a}.
    \item \emph{Egress sniffing:} To observe both ends of the communication, an
		attacker must also observe egress DNS traffic.  We expect the adversary
		either to be on the path between exit relay
		and a DNS server or to run a malicious DNS
		resolver or server.  An attacker may also run an exit relay,
		but in this case conventional end-to-end correlation
		attacks~\cite{Murdoch2007a} are at least as effective as those we
		describe here.
\end{itemize}
We combine a conventional website fingerprinting attack operating on traffic
from ingress sniffing with
DNS traffic observed by egress sniffing, creating \name attacks. Our attacks
correlate the web\emph{sites} observed by the website fingerprinting attack in
ingress traffic with
the web\emph{sites} identified from DNS traffic. Next, we describe how we
simulate the DNS traffic from Tor exits, how we map DNS requests to websites,
and finally present our two \name attacks.

\subsection{Approximating DNS traffic from Tor exits}
\label{sec:attack:sim}

We first investigate the type and volume of DNS traffic that Tor's exit relays
send.  There are no logs of outgoing traffic
from Tor exit relays available to us, and ethical considerations kept us
from trying to collect them (\eg, by operating exit relays and recording
all the outgoing traffic). We therefore opt to approximate the DNS traffic
emerging from Tor exit relays by \first building a model of typical Tor
users' website browsing patterns, \second collecting a minimally invasive
dataset of DNS traffic, and \third accounting for the effects of DNS caching.

\subsubsection{Modeling which sites Tor users visit}
\label{sec:attack:pop}

We first build a model to approximate {which websites} Tor users visit.
As of July 2016, there are about 173 million active
websites~\cite{numberofwebsites}; the Alexa ranking~\cite{alexatop1k}
gives insights into their popularity based on the browsing behavior of
a sample of all Internet users. The distribution of the popularity of
these websites has previously been fit to a power-law distribution based
on the rank of the
website~\cite{Mahanti2013a,Clauset2009a,Ali2007a}.
For the pageview numbers of the Alexa top 10,000 websites, we found a
power-law distribution to be a good fit as neither a log-normal nor a
power-law distribution with exponential cutoff (\ie, a truncated power-law
distribution) offered significantly better fits.
We used the Python {\tt powerlaw} package~\cite{Alstott2014a} for fitting and
picked a power-law distribution with an $\alpha$ parameter of about $1.13$.
When varying the fitting parameter $x_{min}$ that determines beyond which
minimum value the power-law behavior should hold in the provided data, we can
get different $\alpha$ values. We made a conservative choice of picking this
smaller $\alpha$ value as it underestimates the popularity of popular websites
and therefore is worse for the attacker.\footnote{Alexa's page-view numbers
ignore multiple visits by the same user on the same day (see
\url{https://support.alexa.com/hc/en-us/articles/200449744}), so the ranking
might be slightly off when modeling website visit patterns.}
Thus, we use a power-law distribution to model what websites Tor users visit.
This might overestimate the popularity of higher-ranked websites such as
Facebook and YouTube because we believe that Tor users---who tend to be
privacy-conscious---are more likely to seek out alternatives than the typical
Internet user.  We will discuss the implications of our model for browsing
behavior later.

\subsubsection{Modeling how often Tor users visit each site}
\label{sec:load-freq}
\begin{figure}[t]
	\centering
	\includegraphics[width=\linewidth]{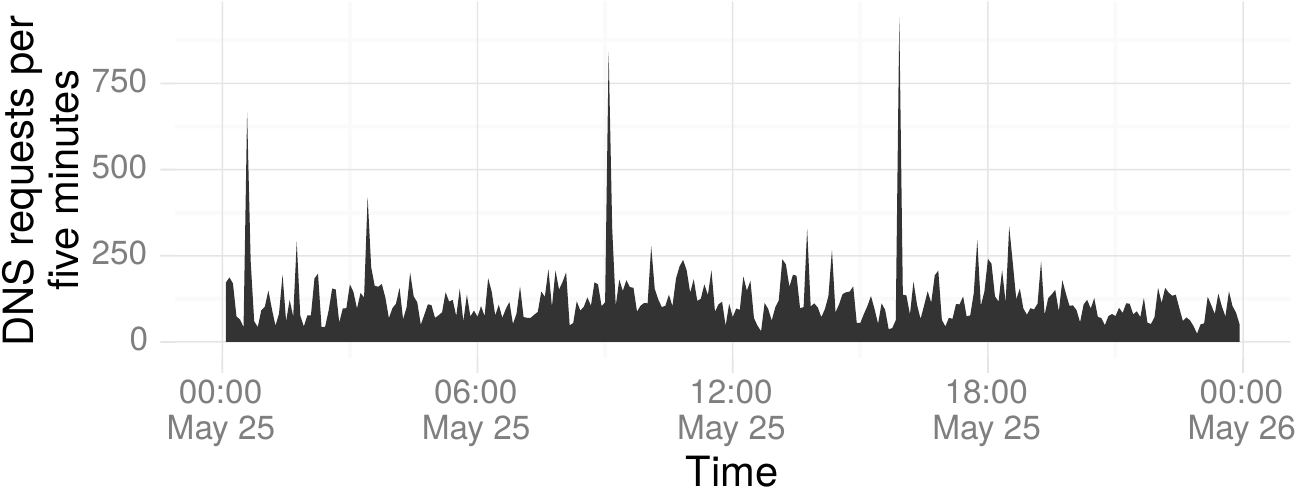}
	\caption{The number of DNS requests per five-minute interval on our
	exit relay for May 25, 2016.  Using a privacy-preserving measurement method,
    we only determined approximate timestamps and no content.}
	\label{fig:dns-reqs}
\end{figure}

Next, we determine how many websites Tor users visit in a certain time span.
We approximated this number by setting up an exit relay whose exit policy
included only ports 80 and 443, so our relay would only forward web traffic.  We
then used the tool {\tt tshark} to capture the timestamps of DNS requests---but
no DNS responses.  We made sure that our {\tt tshark} filter did not capture
packet payloads or headers, so we were unable to learn what websites Tor users
were visiting.  In addition, we patched {\tt tshark} to log timestamps at a
five-minute granularity. The coarse timing granularity allows us to publish this
dataset with minimal privacy implications; Section~\ref{sec:ethics} discusses
the ethical implications of this experiment in more detail.  We ran the
experiment for approximately two weeks from May 15, 2016 to May 31, 2016, which
allowed us to determine the number of DNS requests for 4,832 five-minute
intervals.  Figure~\ref{fig:dns-reqs} shows this time series, but for clarity we
only plot May 25, 2016.  The distribution's median is 105.
The time series features several spikes; the most
significant one counts 1,410 DNS requests.  We repeated the same experiment with
the so-called \emph{reduced exit policy}\footnote{The reduced exit policy is
	available online at
\url{https://trac.torproject.org/projects/tor/wiki/doc/ReducedExitPolicy}.}
because it contains several dozen more ports and it is more popular among Tor
relay operators; as of August 2016, it is used by 7.8\% of exit relays by
capacity.  In comparison, the exit policy containing only port 80 and 443 only
accounts for 1.5\%.  The reduced exit policy resulted in a median of 102 DNS
requests per five minutes, so the difference between both policies is only three
DNS requests.

We then interpolate these numbers to all Tor exit relays based on their
published bandwidth statistics.  While we measured a median of 105,
the mean of the distribution was 119.3 per five minutes during a two-week period.
From DNS statistics of the Alexa top one million websites (see
Section~\ref{sec:dns2site}) we know that one website visit causes outgoing DNS requests for 10.3 domains on average
(assuming a power-law distribution of site popularity as described above, and
taking into account Tor's caching of pending DNS requests, ensuring that multiple
requests sent by clients for the same domain name only result in one outgoing request
by the exit).
This means that we had an average of about 23.2 website visits per ten
minutes on our exit relay.  Assuming that the two main factors influencing the
volume of DNS requests are a relay's \emph{bandwidth} and \emph{exit policy},
and having shown that the exit policy does not significantly impact
the number of DNS requests, we can scale this number up to the whole Tor network
using the provided bandwidth of exit relays.  In particular, we use the
self-reported bandwidth information from Tor exit
relays collected in the so-called extra-info descriptors available on
CollecTor~\cite{collector} and estimate the number of website visits on
each of the about 1,200 exit relays active at that time. The resulting average
total number of websites visited through the Tor network is about 288,000
websites per ten minutes. This number can, however, only be considered
to be a rough estimate, as the interpolation is based on only one exit
relay and the bandwidth data of the other exit relays is self-reported
and might therefore be incorrect or incomplete. For instance, the result
might under-estimate the size of the Tor network if not all exit relays
published their complete bandwidth statistics for the considered
two-week period.

Recently, Jansen and Johnson~\cite{Jansen2016a} measured that the average number
of active web (port 80 and 443) circuits in Tor amounts to about 700,000 per ten
minutes.\footnote{This number comes from a preprint kindly
shared by the authors and might change in the final version.}
Tor Browser, The Tor Project's fork of Firefox, builds one circuit per
website entered in the URL bar. How long the circuit remains active depends on
Tor Browser settings (primarily {\tt MaxCircuitDirtiness} currently set to ten
minutes) and how long TCP streams in the circuit are active: as long as at
least one stream is active, the circuit remains active.  Each time a new stream
is attached to a circuit, the circuit's dirtiness timeout is reset.
The number of active circuits serves as an
upper bound for the number of websites visited over Tor: visiting different
pages of a website will use the same circuit, and visiting a new website will
construct a new circuit.  Users visiting several pages of a website and websites
with long-lived reoccuring connections, like Twitter and Facebook with
continuously updating feeds,
all lower the number of websites visited in Tor relative to the number of active
circuits. 
For our model we consider the upper bound of 700,000 to be the number
of websites visited through the Tor network per ten minutes. This is a
conservative choice as more website visits increase the anonymity set of
websites possibly visited by a Tor user---and therefore reduces the
information an attacker can gain from observed DNS data. 
In later sections we revisit the implications of our choice by both
scaling the Tor network up to ten-times its estimated size, and scaling
it down to the size of 288,000 website visits per ten minutes that we
got from our own interpolation described above. 

\subsubsection{Modeling the effects of DNS caching at Tor exits}
To analyze which DNS requests the adversary can see, we need to
take caching of DNS responses into account. We ignore client-side DNS
caching since it is disabled by default, as described in
Section~\ref{sec:background}.
On the exit relays, which perform DNS resolution on behalf of Tor clients,
caching is relevant because all Tor clients using the same exit relay share its
cache.  An exit relay maintains its own DNS cache%
\footnote{The code is available online at \url{https://gitweb.torproject.org/tor.git/tree/src/or/dns.c?id=tor-0.2.9.1-alpha}.}
(in addition to the cache of its DNS resolver) and enforces a minimum TTL of 60
seconds and a maximum TTL of 30 minutes.%
\footnote{The code is available online at \url{https://gitweb.torproject.org/tor.git/tree/src/or/dns.c?id=tor-0.2.9.1-alpha\#n209}.}
We refer to this as Tor's \emph{TTL clipping}. However, due to a bug in the
source code that we identified,%
\footnote{The bug report is available online at \url{https://bugs.torproject.org/19025}.}
the TTL of all DNS responses is set to 60 seconds.

If a Tor client attempts to resolve a domain that an exit relay already has in
its cache, the adversary will be unable to observe this request.  However, the
adversary can record all observed DNS requests over the past $x$
seconds, where $x$ is the maximum TTL value (\ie, maintain a sliding window of
length $x$).  If a Tor client
is attempting to resolve a domain name, the request is either cached or not.  If
it is not cached, the adversary will see it as a new, outgoing DNS request from
the exit relay. If it is cached, it must have been resolved by the exit relay in
the last $x$ seconds, and will therefore be in the sliding window.  The sliding
window technique allows the attacker to observe all DNS requests, regardless of
if they are cached or not at Tor exits.

We assume that an adversary applies this sliding window technique and models the
observable DNS data accordingly.  The attacker observes a fraction of Tor's exit
bandwidth for a specific window length, and together with our website visit
frequency estimation, this triggers a number of website visits in our
simulation.  For each visit event, we randomly draw a website using the
power-law website popularity distribution described above and put its DNS
requests into the window. As we will see next, we do not need to simulate or
consider the fact that the observed fraction of Tor exit bandwidth corresponds
to many different exits with individual caches.

\subsection{Inferring website visits from DNS requests}
\label{sec:dns2site}

Given a sliding window of DNS requests, we investigate how this information can
help determine whether a user has visited a website of interest.  In April 2016,
we visited the Alexa top one million websites five times, and collected all DNS
requests that each visit of a website's frontpage generated.  We refer to the
data collected for one visit as a \emph{sample}.  We performed these
measurements in five rounds from a university network, where each round browsed
all one million websites in a random order before visiting the same website
again. We used Tor Browser~5.5.4 and configured it {\em not to browse over Tor}:
Tor Browser ensures that the browser behavior is identical to a Tor Browser user
over Tor. By not using Tor, we can bypass IP blacklists and CAPTCHAs that are
triggered by IP addresses of exit relays~\cite{Khattak2016a}.
Table~\ref{tab:dns-censor} shows the percentage of websites in our dataset that
are hosted by CloudFlare or Akamai.  We might not be able to access these
websites programatically over Tor as they block or filter exit relays, as
identified by Khattak \ea~\cite{Khattak2016a}. We also include Google, which is
prevalent in our dataset and restricts access to Tor users for Google's search.

\begin{table}[t]
	\renewcommand{\tabcaptext}{The percentage of websites on Alexa top-1 million websites using providers
	involved in censoring or restricting access from
        Tor~\protect\cite{Khattak2016a}.}
      \topcap{\tabcaptext}
	\centering
	\begin{tabular}{l r}
	\toprule
	\textbf{Description} & \textbf{Percentage} \\
	\midrule
	Website behind CloudFlare IP & 6.44 \\
	Domain on website uses CloudFlare & 25.81 \\
	Domain on website uses Akamai & 33.86 \\
	Domain on website uses Google & 77.43 \\
	\bottomrule
	\end{tabular}
        \bottomcap{\tabcaptext}
	\label{tab:dns-censor}
\end{table}

\begin{figure}[t]
	\centering
	\includegraphics[width=0.75\linewidth]{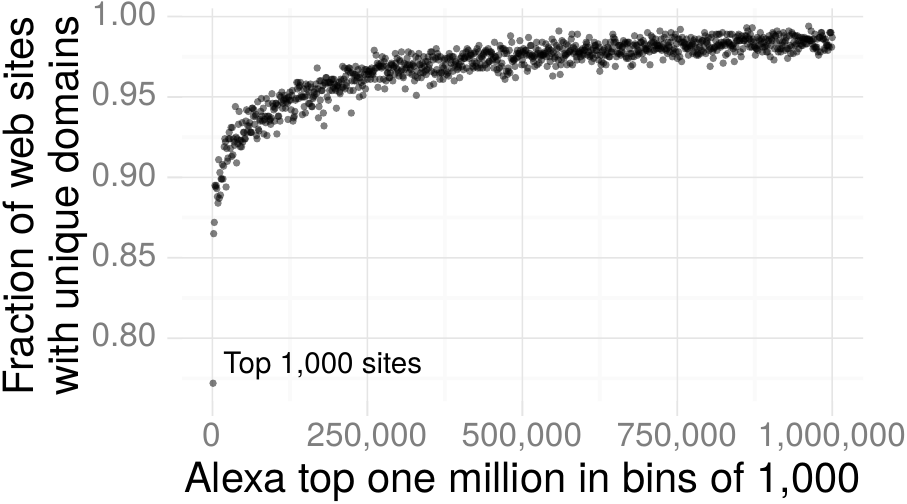}
	\caption{The fraction of websites in Alexa's top one million that have at
	least one unique domain.  We grouped all domains into 1,000 consecutive,
	non-overlapping bins of size 1,000.  The vast majority of sites (96.8\%)
	have unique domains.}
	\label{fig:unique-domains}
\end{figure}

We collected 2,540,941 unique domain names from a total of 60,828,453 DNS
requests. The dataset contains 2,260,534 domains that are unique to a particular
website, \ie, are not embedded on any other top million site; we call these
domains {\em unique domains}. Unique domains are particularly interesting
because they reveal to the adversary what sites among the top million the user
has visited.  This is not possible for domains such as {\tt youtube.com}, simply
because many websites embed YouTube videos.  Figure~\ref{fig:unique-domains}
shows the fraction of sites with unique domains for websites up to Alexa's top
one million.  We grouped all domains into 1,000 consecutive, non-overlapping
bins of size 1,000.  For 96.8\% of all sites on the Alexa top one million there
exists at least one unique domain.  Interestingly, more popular websites are
less likely to have a unique domain associated with them: only 77\% of the first
bin---the most popular 1,000 domains---contain at least one unique domain,
significantly less than the rest of the data.

Table~\ref{tab:dns-domains} shows summary statistics for the number of domains
per website. At least half of the sites have ten domains per website, two of
them are unique, suggesting that an adversary can identify many website visits
by observing a single unique DNS request.

\begin{table}[t]
	\renewcommand{\tabcaptext}{Summary statistics for the number of domains per
	website in the Alexa top 1 million. More than half of the sites embed two
	domains that are unique to that site.}
	\topcap{\tabcaptext}
	\centering
	\begin{tabular}{l r r r r}
	\toprule
	\textbf{Domains} & \textbf{Median} & \textbf{Mean $\pm$ Stddev} & \textbf{Min.} & \textbf{Max.} \\
	\midrule
	Per site & 10 & $12.2\pm11.2$ & 1 & 397 \\
	Unique per site & 2 & $2.3\pm\phantom{0}1.8$ & 0 & 363 \\
	\bottomrule
	\end{tabular}
        \bottomcap{\tabcaptext}
	\label{tab:dns-domains}
\end{table}

To evaluate the feasibility of mapping DNS requests to websites, we
construct a na\"{\i}ve website classifier that maps the unique domains
in a set of DNS requests to the corresponding website that contains a
matching set of domains.  With five-fold cross-validation on our Alexa
top one million dataset (with five samples per site),
we consider a closed world and an open world.
In the closed world, the attacker can use samples
from all sites in training; in the open world, some sites
are unmonitored and therefore unknown (as per the fold).  The
closed-world evaluation yields 0.955 recall.  In the open-world
evaluation, we monitor the Alexa top 500,000 with five samples each and
consider 433,000 unmonitored sites.  The number of unmonitored sites is
determined by our power-law
distribution to represent a realistic base rate (for the entire Tor network)
for evaluating our classifier: on average, for sites on Alexa top 500,000
to be visited 2.5 million times there will be about 433,000 visits to sites
outside of Alexa top 500,000.  Our classifier does not take into account the
popularity of websites.
The open-world evaluation yields a
recall of 0.947 for a precision of 0.984.  By accounting for the order
of requests, per-exit partitioning of DNS requests, TTLs, and website
popularity, we expect that classifying website visits from DNS requests
might be made even more accurate.
Further, a closed world is a realistic setting:
determining the DNS requests made by all 173 million active websites on the
Internet is practical, even with modest resources.
We use the conservative open world results when simulating the Tor network and
the attacker's success in mapping DNS requests to websites.
We conclude from our results that observing DNS requests coming out of Tor is
almost as effective at identifying websites as observing the web traffic
itself.

\subsection{Classifiers for \name attacks}

We use Wa-kNN from Wang \ea~\cite{Wang2014a} (described in
Section~\ref{sec:background}) and a list of sites derived from
observing DNS requests to implement two \name attacks:

\begin{description}
	\item[\texttt{ctw}] We ``close the world''
	on a Wa-kNN classifier that we modified to consider only the distance to
	observed sites when calculating the $k$-nearest neighbors.
	The classifier still considers the distance to all unmonitored sites.
	\item[\texttt{hp}] When Wa-kNN classifies a trace as a monitored site, confirm
	that we observed the same site in the DNS data (ensuring {\em high
	precision}). If not, make the final classification unmonitored.
\end{description}
\noindent
These approaches apply to any website fingerprinting attack. The
\texttt{ctw} attack increases the effectiveness of conventional website
fingerprinting attacks by making them more akin to a closed-world setting,
where websites have known fingerprints and often the world is of limited size.
Conceptually, the attack could also include
a custom weight-learning run---training only on observed sites---but our initial
results noted little to no gain, despite significant increases in
testing time.
We expect that this is due to the fact that some features of traffic traces are
more useful than others, regardless of the training data~\cite{Hayes2016a}.
The \texttt{hp} attack only produces a positive classification if both ingress
and egress traffic are consistent, resulting in a simple but effective
classifier.

\section{Evaluating \name Attacks}
\label{sec:analysis}

\subsection{Attack precision and recall}

To evaluate our \name attacks, we collected traffic traces in May
2016 using Tor Browser~5.5.4.
We modified Tor Browser to not generate network traffic on launch
(\ie, check for
updates, extensions, \etc), and we modified
Tor (bundled with Tor Browser) to log incoming and outgoing cells.
We then performed 100 downloads for each site in the Alexa top
1k and one download for each site in the
Alexa top (1k,101k]. We randomly distributed these measurement tasks
to a Docker fleet; each download used a fresh circuit without
guard relay, and a fresh copy of Tor Browser for up to 60 seconds,
in line with the recommendations by Wang and Goldberg~\cite[\S~4]{Wang2013a}.
We cached Tor's network consensus to avoid causing excessive load on the
network.  We considered the measurement to be successful if we managed to
resolve the domain of the site; we did not prune our dataset further, neglecting
issues like CloudFlare CAPTCHAs, outliers, control cells, and localized
domains~\cite{Juarez2014a}.  Presumably, this means that we will underestimate
the effectiveness of our attack, but we are primarily interested in the
difference between website fingerprinting attacks and \name
attacks~\cite{Wang2013a}.

We perform ten-fold cross-validation for all of our experiments in the open
world setting, monitoring 1,000 sites with 100 instances each, and
100,000 unmonitored sites.
Note the 1:1 ratio between monitored traces and unmonitored traces,
ensuring that for the classifier there is equal probability in the testing
phase that a trace is a monitored or unmonitored site.
In other words, the \emph{base rate} is 0.5 in our experiments.
Furthermore, for all experiments we specify the starting Alexa rank of the
monitored sites
\emph{when simulating sites visited over the Tor network}.
We always use the same sample data for website fingerprinting.
When monitoring 1,000 sites starting at rank 1, sites
[1,1000] are monitored and Alexa [1001,101000] unmonitored. Starting to
monitor Alexa from rank 100 means that Alexa sites [101,1100] are monitored,
and Alexa [1,100] and [1101,10100] unmonitored.
We never monitor an unmonitored site or vice versa.
How popular monitored sites
are is a key factor in the effectiveness of our attacks.

\begin{figure}[t]
\centering
\includegraphics[width=0.465\linewidth]{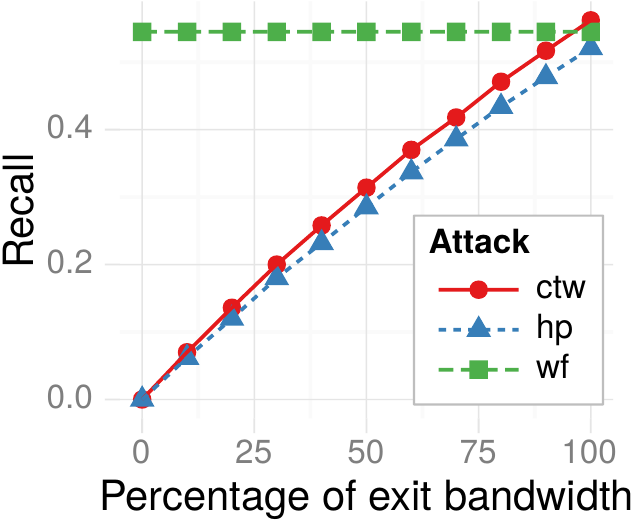}
\includegraphics[width=0.465\linewidth]{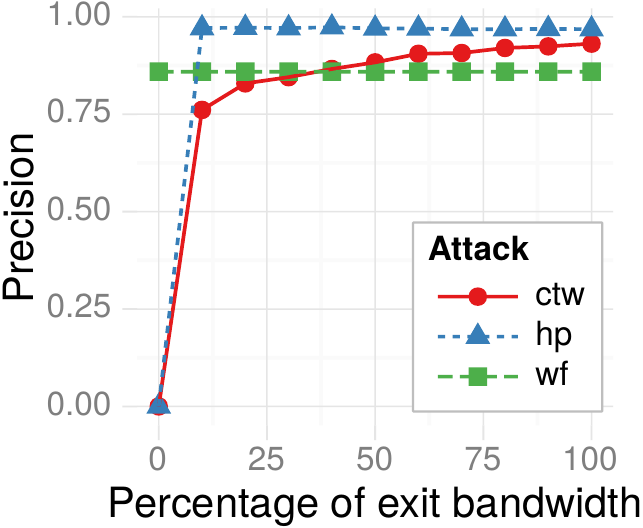}
\caption{Recall and precision for an open-world dataset with monitored sites
at Alexa rank 10k and lower. We compare our \name attacks (\texttt{ctw and
 \texttt{hp}}) to a conventional website fingerprinting attack (\texttt{wf}) for different
 percentages of observed exit bandwidth. }
\label{fig:fpt:torpct}
\end{figure}

Figure~\ref{fig:fpt:torpct} shows the recall and precision for our \name
attacks as a function of the percentage of observed Tor exit bandwidth by the
attacker monitoring Alexa sites for sites whose ranks are 10,000 or less.
For recall, both \texttt{ctw} and \texttt{hp} are bound by the
percentage of exit bandwidth observed by the attacker (the percentage is an
upper bound).
It is simply not possible to identify a monitored site in the DNS data that
the attacker does not see. At 100\% of exit bandwidth, \texttt{ctw} sees
better recall than \texttt{wf}. For \texttt{hp} the results suggest that:
\begin{equation}
	\label{eq:hprecall}
	\textnormal{recall}_{\texttt{hp}} = \textnormal{recall}_{\texttt{wf}} * \textnormal{pct}
\end{equation}
The above relationship only holds when observing DNS requests gives
a clear advantage to \texttt{hp} in terms of precision over \texttt{wf} (see
the following paragraph).
For precision the \texttt{hp} attack has an immediate gain over \texttt{wf} as
soon as the attacker can observe {any exit bandwidth}.
Although the \texttt{hp} attack has near-perfect precision, the
\texttt{ctw} attack benefits from observing increasingly more exit traffic,
nearly reaching the same levels as \texttt{hp} at 100\% of the exit bandwidth.

\begin{figure}[t]
\centering
\includegraphics[width=0.465\linewidth]{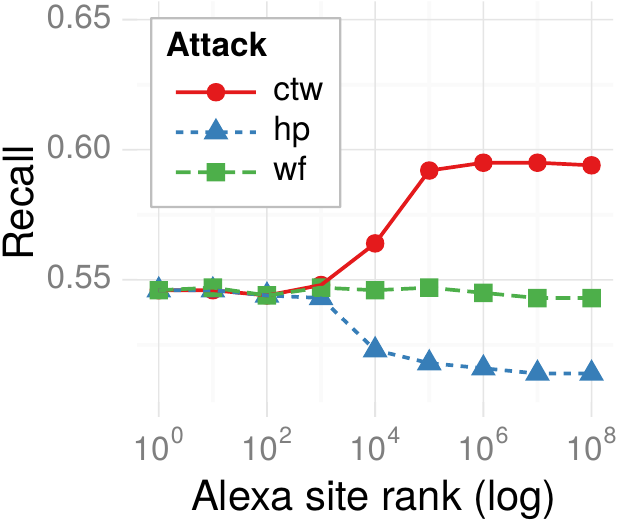}
\includegraphics[width=0.465\linewidth]{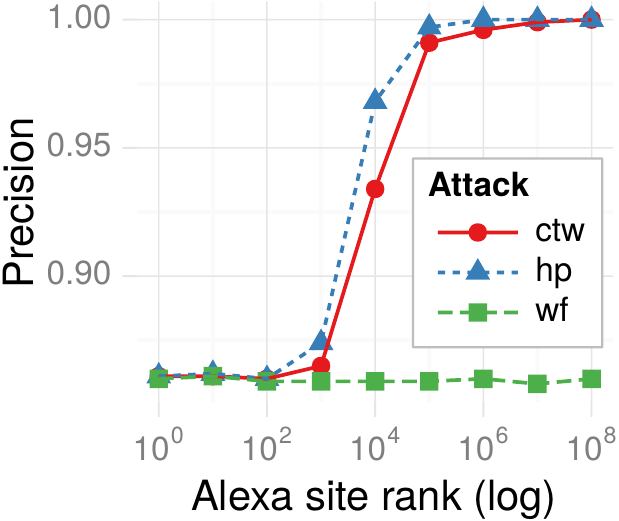}
\caption{The recall and precision when varying the starting Alexa rank of
monitored sites for 100 percentage of exit bandwidth.}
\label{fig:fpt:alexa}
\end{figure}

Figure~\ref{fig:fpt:alexa} shows recall and precision at 100\% of
observed Tor exit bandwidth as a function of the starting Alexa rank of
monitored sites (we still monitor 1,000 sites).
For popular websites (\ie, websites with a high Alexa ranking),
there is no difference between our attacks and the
\texttt{wf} attack. This is because even with a window of only 60 seconds,
it is almost certain that at least one user visited any of the most popular
sites over Tor. For sites that rank 1,000 or lower (\ie, less popular sites),
both \name attacks show a clear improvement in precision while
\texttt{ctw} also shows improved recall---but only at 100\% observed exit
bandwidth, as shown in Figure~\ref{fig:fpt:torpct}.
These results paint a bleak picture: a small fraction of exit
bandwidth provides a perfectly precise attack on relatively
{unpopular} sites such as wikileaks.org, which had Alexa rank
10,808 on April 15th 2016.

\subsection{Sensitivity analysis}

\begin{figure}[t]
\centering
\subfigure[Estimating the effect of website fingerprinting defenses.]{
	\includegraphics[width=0.465\linewidth]{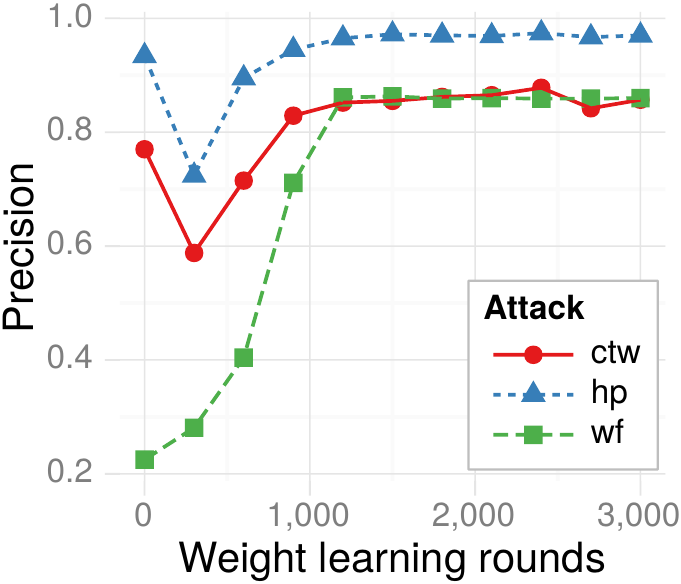}
    \label{fig:fpt:var:rounds}
}
\subfigure[Effect of increasing the analysis time window due to TTL clipping.]{
	\includegraphics[width=0.465\linewidth]{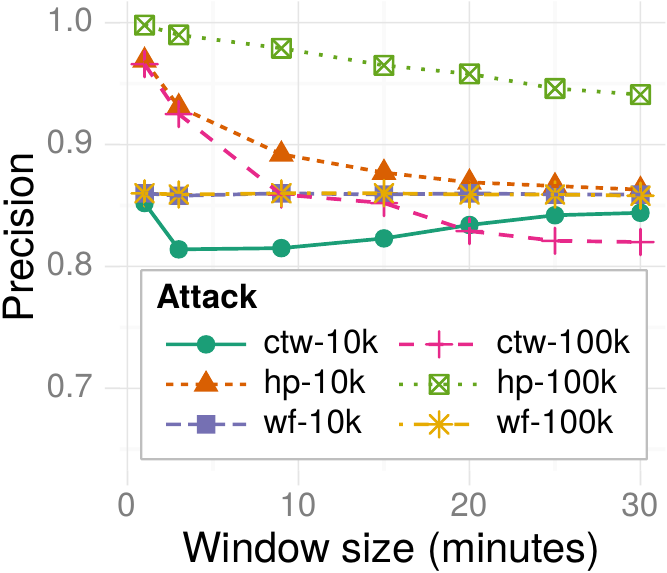}
    \label{fig:fpt:var:window}
}
\subfigure[Effect of Tor network scale for Alexa ranks 10k and 100k.]{
	\includegraphics[width=0.465\linewidth]{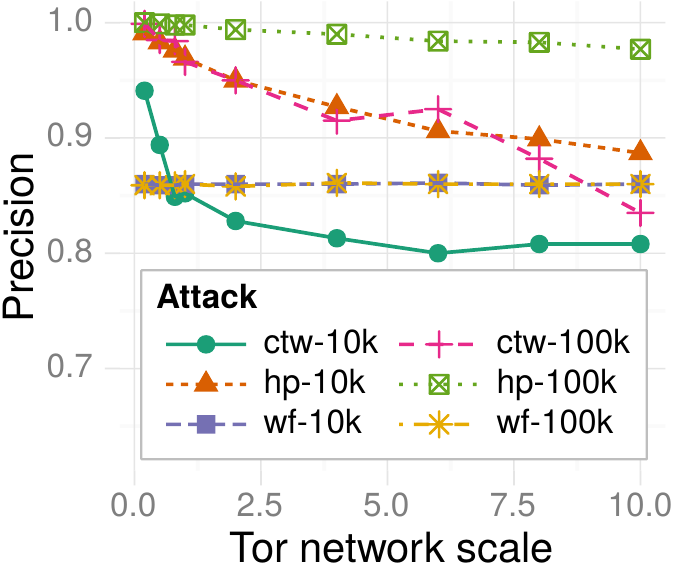}
    \label{fig:fpt:var:scale}
}
\subfigure[Effect of different website popularity distributions.]{
	\includegraphics[width=0.465\linewidth]{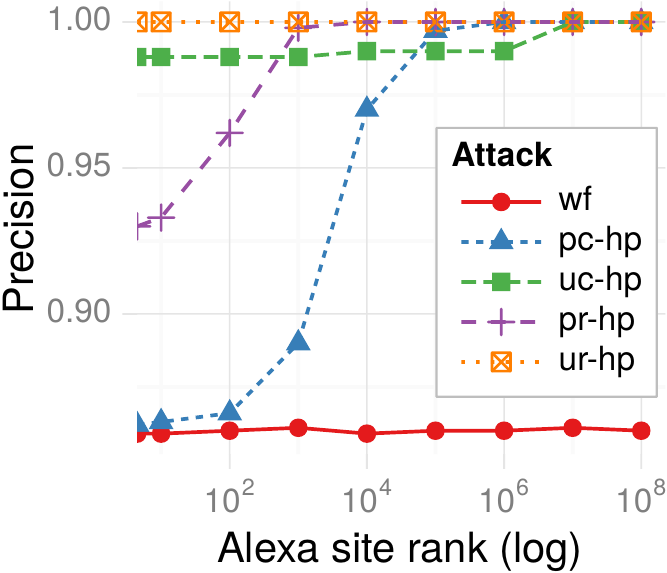}
    \label{fig:fpt:var:dist}
}
\caption{The effect on attack precision. The defaults are: Alexa from top 10,000,
2,500 weight-learning rounds,
60-second window size, Tor network scale 1.0, and the conservative
power-law distribution (pc) with $\alpha=1.13$.}
\label{fig:fpt:var}
\end{figure}

To better understand the extent and limitations of our attacks, we
study the sensitivity of our results to website fingerprinting defenses,
TTL clipping, the growth of the Tor network, and website popularity
distribution.  In this section, we assume that an adversary can observe Tor
exit relays representing 33\% of exit bandwidth (as observed on average
by Google) and consider only precision (where we see clear gain from both our
attacks).  Unless stated otherwise, we perform our evaluation on
websites starting at Alexa rank 10,000 and proceeding through less
popular websites, use 2,500 weight-learning rounds, have a 60-second window
size, a Tor network scale of 1.0, and use the conservative power-law
distribution from Section~\ref{sec:attack:pop}.

\subsubsection{Effect of website fingerprinting defenses}

Website fingerprinting defenses are being
designed and discussed for deployment in Tor~\cite{adapativepadding}.
The defenses produce bandwidth and/or latency overhead, with a significant
increase in overhead as the defense becomes stronger.
For example, Juarez \ea
observe an exponential increase in bandwidth overhead as the protection of the
defense WTF-PAD~\cite{Juarez2016a} increases.
This leads to the goal of finding an optimum that provides strong protection
while also keeping the overhead tolerable for Tor users.
As a first attempt to approximate the effect of fingerprinting
defenses on \name attacks, we use Wa-kNN with
random weights and no weight-learning: this significantly reduces the
effectiveness of the attack since some features (like indices of outgoing
packets) are several orders of magnitude more useful
than others~\cite{Juarez2016a}.

Figure~\ref{fig:fpt:var:rounds} shows the effects of between 0 and 3,000
rounds of weight-learning. At few to no rounds, the precision for the
\texttt{wf} attack is below 50\%---the classification is more likely to be wrong
than right---while there is no impact on the \texttt{hp} attack and a relatively
small decrease for the \texttt{ctw} attack.
For recall (not shown in the figure), the bound and relationship is
as in Equation~\ref{eq:hprecall}: for \texttt{wf}, at zero rounds, recall is
0.055; for \texttt{hp} at zero rounds, recall is 0.019. These results suggest
that for website fingerprinting defenses to be effective against \name attacks,
the defense must be tuned to provide a low recall even when the parameters of
website fingerprinting attacks are selected for high recall.

\subsubsection{Effect of Tor's TTL clipping}

\begin{table}[t]
\renewcommand{\tabcaptext}{Median and mean DNS TTL values across Alexa top one million sites. Raw
TTLs are unprocessed, as they appear in DNS lookup traces. Tor TTLs adhere to
Tor's TTL clipping.  Unique refers to the TTLs for unique domains; min unique
only considers the unique domains with the minimum TTL for each
website.}
\topcap{\tabcaptext}
\centering
\begin{tabular}{l r r}
\toprule
\textbf{TTLs} & \textbf{Median TTL (sec)} & \textbf{Mean TTL (sec) $\pm$ Stddev} \\
\midrule
Raw & \multirow{ 2}{*}{255} & $9{,}780.0\pm42{,}930.5$ \\ 
Tor &  & $701.5\pm\phantom{00{,}}755.3$ \\ 
Unique raw & \multirow{ 2}{*}{900} & $13{,}022.2\pm35{,}054.4$ \\ 
Unique Tor &  & $1{,}005.3\pm\phantom{00{,}}789.6$ \\ 
Min unique raw & \multirow{ 2}{*}{60} & $3{,}833.9\pm11{,}073.6$ \\ 
Min unique Tor & & $644.2\pm\phantom{00{,}}763.8$ \\ 
\bottomrule
\end{tabular}
\bottomcap{\tabcaptext}
\label{tab:ttls}
\end{table}

As discussed in Section~\ref{sec:attack:sim}, due to a bug in Tor, all exit
relays cache DNS responses for 60 seconds, regardless of the DNS response's TTL.
Therefore, a sliding window covering the last 60 seconds of all observed DNS
requests suffices to capture all monitored sites through Tor (subject to the
fraction of observed Tor exit bandwidth, and mapping DNS requests to sites).

Table~\ref{tab:ttls} shows the TTL of DNS records in our Alexa top one million
sites dataset from Section~\ref{sec:dns2site} for the
TTL as-is (raw) and when clipped by Tor.
(We emulate what the correct TTL values
would be due to TTL clipping supposing that Tor eventually fixes the
aforementioned bug.)  For each of these cases, we
also consider TTLs for all unique domains, and for only the unique
domain for each website with the lowest TTL.  About half of the sites on
Alexa top one million has a unique domain with a TTL of 60 seconds or
less; 48\% of the raw unique TTLs are below 60 seconds and only 26\%
above 30 minutes. Fixing the Tor clipping bug is therefore not
sufficient; to mitigate \name attacks, the minimum TTL should be
significantly increased.  In this case, we find that Tor's TTL clipping has
no effect on the median TTL, but significantly reduces the mean TTL.

Suppose that Tor eventually fixes the DNS TTL bug, requiring the
attacker to monitor DNS lookups for a time interval equal to the maximum
TTL of all unique domains for any monitored site.
Figure~\ref{fig:fpt:var:window} shows the effect on precision for
different time intervals from 60 seconds to 30 minutes (Tor's {\tt
  MAX\_DNS\_ENTRY\_AGE} for keeping entries in an exit's DNS resolver
cache), and for Alexa starting rank 10,000 and 100,000. For \texttt{ctw},
the time interval has a significant effect on both Alexa starting ranks,
while \texttt{hp} is only affected for sites ranked 10,000 or higher;
for less popular sites, the DNS lookup data still significantly improves
fingerprinting precision, even with the larger window size.

\subsubsection{Effect of Tor network growth}
Figure~\ref{fig:fpt:var:scale} scales the size of the Tor network with
respect to site
visits from the status quo to ten times its size, for Alexa starting rank 10,000 and
100,000. At twice its current size, the impact on \name attacks is smaller than
increasing the minimum TTL for DNS caching to three minutes as shown in
Figure~\ref{fig:fpt:var:window}. These results indicate that \name
attacks will remain
practical for many sites in the Alexa top one million, even as the Tor network grows.


\subsubsection{Sensitivity to website popularity distribution}

To explore the sensitivity of our results to different distributions in
how users visit websites, we now evaluate the effectiveness of \name
attacks with four different website distributions:
\begin{description}
	\item[pc] A conservative power-law distribution
	(with $\alpha=1.13$)
	that we manually fitted to the Alexa top 10,000 data,
	slightly underrepresenting the popularity of top Alexa sites.
	We described this distribution in Section~\ref{sec:attack:pop}.
	\item[pr] A realistic power-law distribution
	(with $\alpha=1.98$)
	that is the best fit according to
	the Python {\tt powerlaw} library by Alstott~\ea~\cite{Alstott2014a} for the Alexa
	top 10,000 data.
	\item[uc] A conservative uniformly random distribution that
	only considers one million active websites browsed over Tor.
	\item[ur] A realistic uniformly random distribution that
          considers 173 million active websites, as reported by Netcraft
          in July 2016 for the Internet~\cite{numberofwebsites}.
\end{description}
Figure~\ref{fig:fpt:var:dist} shows the effect on the precision of the
\texttt{hp} attack for the different distributions as we vary the starting
Alexa rank. The uniform distributions always have nearly perfect precision.
The difference between the two power-law distributions is about one order of
magnitude in terms of starting Alexa rank: the realistic distribution gets
near perfect at 1,000 and the conservative at 10,000.
We conclude that \name attacks are perfectly precise for unpopular sites
where the probability that someone other than the target browses to a monitored
site within the timeframe given by the window length is negligible.

\section{Internet-scale analysis}
\label{sec:internet-scale}

In the preceding sections we have presented our \name attacks and evaluated
their effectiveness, but we have yet to understand what entities can mount them.
In this section, we aim to quantify the likelihood that any AS is in a position
to mount \name attacks.

\subsection{Approach}

\begin{figure}[t]
	\centering
	\includegraphics[width=\linewidth]{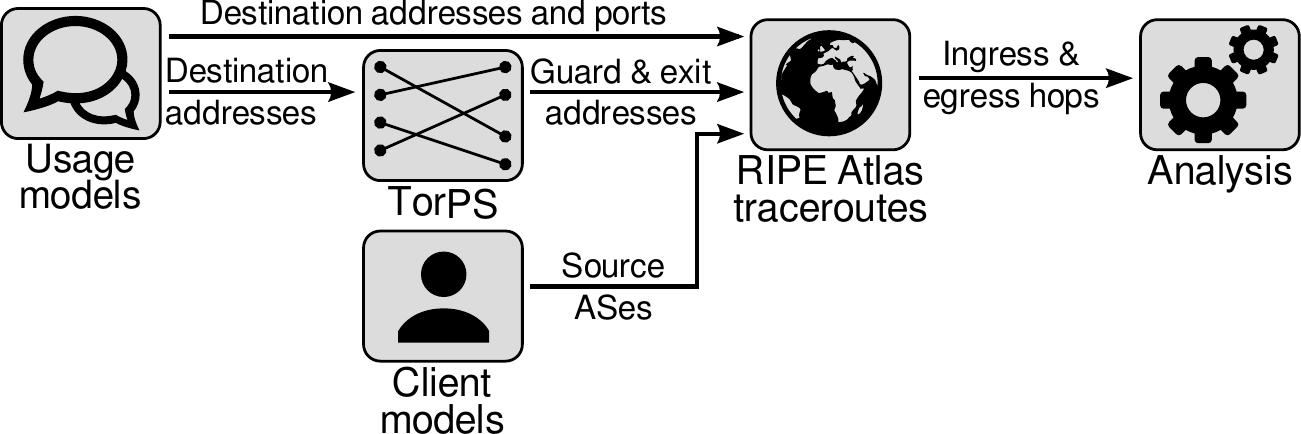}
	\caption{The relation among our simulation components.  Our goal is to
	determine the ASes a Tor user's traffic traverses into and out of the Tor
	network.  Duplicate ASes on both sides can deanonymize streams.}
	\label{fig:simulations}
\end{figure}

Figure~\ref{fig:simulations} summarizes our simulation approach, which we detail
in the next section.  In short, we model the activity of Tor users and simulate
their path selection using TorPS~\cite{TorPS}.  TorPS returns guard and exit
relays, which we then feed as input---together with source ASes and destination
addresses---into our framework that runs traceroutes from RIPE Atlas nodes.  The
rest of this section describes our approach in detail.

\subsubsection{Attack model}

We assume that an AS can mount \name attacks if it can see both traffic
\emph{entering} the Tor network and DNS traffic \emph{exiting} the Tor network.
Recall that an exit relay can perform DNS resolution in two ways; by running a
local resolver, or by relying on a third-party resolver, such as its ISP's or
Google's public resolver.  In the case of exit relays that perform local
resolution, an effective position for an attacker is both \first~anywhere on the
AS path between a Tor client and its guard relay; and \second~anywhere on the
path between an exit relay and any of the name servers the exit has to
communicate with to resolve a domain.  These name servers include the full DNS
delegation path, \ie, a root name server plus subsequent name servers in the DNS
hierarchy.  All ASes along the path from the exit relay to the name servers will
be able to see the domain names that the exit relay is querying.  For exit
relays that rely on third-party resolvers, the adversary instead has to be on
the path between the exit relay and its DNS resolver.

\subsubsection{Simulating Tor user activity with TorPS}

To measure the likelihood that an AS can be in a position to perform a \name
attack, we use TorPS~\cite{TorPS}---short for Tor Path Simulator---which mimics
how a Tor client constructs circuits (see ``TorPS'' in
Figure~\ref{fig:simulations}).  TorPS takes as input archived Tor network
data~\cite{collector} and usage models, which are sets of IP addresses that Tor
clients talk to---\eg web servers.  Given this input, TorPS then simulates for a
configurable number of ``virtual'' Tor clients the way they would select guard
and exit relays.  TorPS is based on the Tor stable release in version 0.2.4.23.
For each simulated client, TorPS uses one guard; this guard selection expires
after 270 days.  We use TorPS to simulate the behavior of 100,000 Tor clients
for the entire month of March 2016.


We need to place our simulated Tor clients into an AS (see ``Client models'' in
Figure~\ref{fig:simulations}).  We selected clients in major ISPs in the
top-five most popular countries of Tor usage according to Tor
Metrics~\cite{metrics-countries}.  As of August 2016, the top five countries are
the U.S., Russia, Germany, France, and the U.K.  For the U.S., we chose Comcast
(AS~7922); for Russia, Rostelecom (AS~42610); for Germany, Deutsche Telekom
(AS~3320); for France, Orange (AS~3215); and for the U.K., British Telecom
(AS~2856).

Having placed simulated Tor clients into ASes, we now model their activity over
Tor (see ``Usage models'' in Figure~\ref{fig:simulations}).  We model each
client to have visited several websites every day in March 2016.\footnote{We
modeled our client behavior off of the ``Typical'' model that Johnson
\ea~\cite[\S~5.1.2]{Johnson2013a} used.}  At 9 a.m. EST, the client visits {\tt
mail.google.com} and {\tt www.twitter.com}.  At 12 p.m. EST, the client visits
{\tt calendar.google.com} and {\tt docs.google.com}. At 3 p.m. EST, the client
visits {\tt www.facebook.com} and {\tt www.instagram.com}.  Finally, at 6 p.m.
EST, the client visits {\tt www.google.com}, {\tt www.startpage.com}, and {\tt
www.ixquick.com}, and at 6:20 p.m. EST, the client visits {\tt www.google.com},
{\tt www.startpage.com}, and {\tt www.ixquick.com} again. Each of the 100,000
simulated Tor clients thus had $12 \cdot 31 = 372$ opportunities to be
compromised given 31 days and 12 site visits per day.  TorPS provided a new
circuit every ten minutes, regardless of how many distinct connections the
client made to different sites; it did not provide a new circuit for different
websites if the client visited the group of sites within the same ten-minute
window.

For simplicity, we assume that only one DNS request occurs every time a client
visits a site. For example, in our model, at 9 a.m. one DNS request will occur
for {\tt mail.google.com} and one DNS request will occur for {\tt
www.twitter.com}. At 6 p.m. three DNS requests will occur, and at 6:20 p.m.
those same three DNS requests will occur again.   For now, we do not take
embedded requests (\ie for embedded website content such as YouTube videos) or
caching into account.

\subsubsection{Inferring AS-level paths using traceroutes and {\tt pyasn}}

Our Internet-scale analysis also requires learning the AS-level paths from each
client to its guard, and from its exit to the destination (see ``RIPE Atlas
traceroutes'' in Figure~\ref{fig:simulations}).  We decided against the commonly
applied AS path inference because Juen \ea showed that it can be quite
inaccurate~\cite{Juen2015a}.  Traceroutes, in contrast, yield significantly more
accurate paths, but are difficult to run from Tor relays:  Past work involved
asking relay operators to run traceroutes for the
researchers~\cite[\S~4]{Juen2015a}.  This approach yielded traceroutes from
relays representing 26\% of exit bandwidth, but does not scale well.  Instead of
running traceroutes from Tor relays, we leverage the RIPE Atlas~\cite{atlas}
platform, a volunteer-run network measurement platform consisting of thousands
of lightweight and geographically spread \emph{probes} that can be used as
vantage points for traceroutes.  Our key observation is that RIPE Atlas has
probes in many ASes that also have Tor relays.  We leverage this observation by
designing measurements to run traceroutes from Atlas probes that are located in
the same AS as exit relays, to each of the destinations in question.

Table~\ref{tab:atlas-coverage} shows that for a day in May 2016, we found that
RIPE Atlas had probes in 52\% of ASes that contain exit relays, and in 51\% of
ASes that contain Tor guard relays.  More importantly, we found that Atlas ASes
cover 58\% of exit \emph{bandwidth} and 74\% of guard bandwidth.  This statistic
is important given that Tor clients select relays weighted by their bandwidth,
and the bandwidth of Tor relays is not uniformly distributed.  Given the growth
of both Tor and Atlas, we expect these numbers to increase in the future.  In
addition to Atlas, we also considered using PlanetLab~\cite{planetlab} to
initiate traceroutes, but unfortunately most PlanetLab nodes are located in
research and education networks~\cite{Banerjee2004a}, and are thus not
well-suited for performing our measurements.

\begin{table}[t]
	\renewcommand{\tabcaptext}{The coverage of RIPE Atlas nodes that are
	co-located with Tor guard and exit relays as of May 2016.}
	\topcap{\tabcaptext}
	\centering
	\begin{tabular}{l|r r}
	\toprule
	\textbf{Atlas probe coverage} & \textbf{Tor guard ASes} & \textbf{Tor exit ASes} \\
	\midrule
	By bandwidth & 73.59\% & 57.53\% \\
	By number & 50.69\% & 52.25\% \\
	\bottomrule
	\end{tabular}
	\bottomcap{\tabcaptext}
	\label{tab:atlas-coverage}
\end{table}

We performed traceroutes from the five Tor client ASes outlined above to all
their respective guard relay IP addresses that TorPS determined.  To measure the
paths from exit relays to their DNS resolvers, we performed the following
traceroutes, simulating four different DNS configurations:
\begin{itemize}
	\item \emph{ISP DNS:} To investigate the scenario in which an exit relay
		uses its ISP's resolver, we chose to represent this as the resolver
		being in the same AS as the exit relay.  Thus, no traceroutes were
		necessary for this experiment.  We acknowledge that this is not
		necessarily the case, but assume that it holds for the majority of exit
		relays.

	\item \emph{Google DNS:} This scenario represents an exit relay using
		Google's public resolver.  To measure the AS path, we perform
		traceroutes from a RIPE Atlas node in the AS of the exit relay to
		Google's public DNS resolver, \ie, {\tt 8.8.8.8}.

	\item \emph{Local DNS:} To measure the paths that would be traversed if an
		exit relay were running its own, local resolver (\eg, the popular
		service {\tt unbound}), we used the command line tool {\tt dig} with the
		{\tt +trace} option to determine the iterative resolution process.  We
		tracked all name server IP addresses from referrals at each level of the
		delegation path, and performed traceroutes to those IP addresses.

	\item \emph{Status quo:} This scenario represents the state of the Tor
		network as of March 2016, a combination of the above configurations.
		Recall that in Section~\ref{sec:mapping-resolvers}, we determined
		the IP addresses of the resolvers that exit relays use.  We ran
		traceroutes to these very IP addresses.  For the exit relays that used
		several resolvers during March, we randomly assigned one to the relay.
		We ended up having data for 73\% of the exit relays that TorPS ended up
		picking.\footnote{The missing 27\% are due to the churn in exit relays.
		Since we did not run our {\tt exitmap} experiment each hour, we were
		bound to miss some exit relays.}
\end{itemize}

We then mapped each IP address in every traceroute to its corresponding AS (see
``Analysis'' in Figure~\ref{fig:simulations}).  The
Python module {\tt pyasn}~\cite{pyasn} relies on BGP routing tables to perform
these mappings; by using a routing table that coincides with the time when we
performed our traceroutes, we can obtain accurate AS-level mappings.  This
method is subject to inaccuracies due to BGP route hijacks or leaks, but we
expect those events to be relatively unlikely for the time period and IP
prefixes that we are concerned with.

\subsubsection{Putting it all together}
Like Johnson \ea~\cite[\S~4.2]{Johnson2013a}, we consider two security metrics;
we aim to estimate \first the fraction of compromised streams per simulated Tor
user, and \second the amount of time it would take for the first compromise to
occur.  For both metrics, we consider the four DNS configurations outlined
above.  Our simulation can reveal the respective average threat that a given DNS
configuration poses for Tor users.

The traceroutes described in the previous section yielded two sets of ASes, one
from the Tor users' ASes to their guard relays, and one from approximately half
of the exit relays' ASes to the different destinations, which depend on the exit
relays' DNS configurations.  We intersect both AS sets (the ``ingress'' and
``egress'' hops of Figure~\ref{fig:simulations}) and classify a website visit as
compromised if the intersection is non-empty.  As stated earlier, for some exit
relays we did not have associated AS-level paths to a particular destination,
either due to a lack of co-located RIPE Atlas probes, or because of missing
traceroute information.  In these cases, we checked if the exit AS had the
potential to launch an attack, and if not, the stream was considered to be
uncompromised in order to err on the conservative side.

To compute the fraction of compromised streams, we counted the streams that were
compromised for every simulated user out of a possible maximum of 372. To
compute the time until first compromise, we determined the first stream in which
the user was compromised, took its timestamp, and calculated the offset from the
beginning of March 1, 2016.  For users that were not compromised during the
month of March, we assigned a maximum value of 31 days as the time until first
compromise, which is reflected in the plots in our next section.  Users who were
compromised immediately would have a value of 0, signifying that they were
compromised at the very beginning of March 1.

\subsection{Results}

\begin{figure}[t]
\centering
\subfigure[The fraction of compromised streams of simulated Tor clients.]{
	\includegraphics[width=\linewidth]{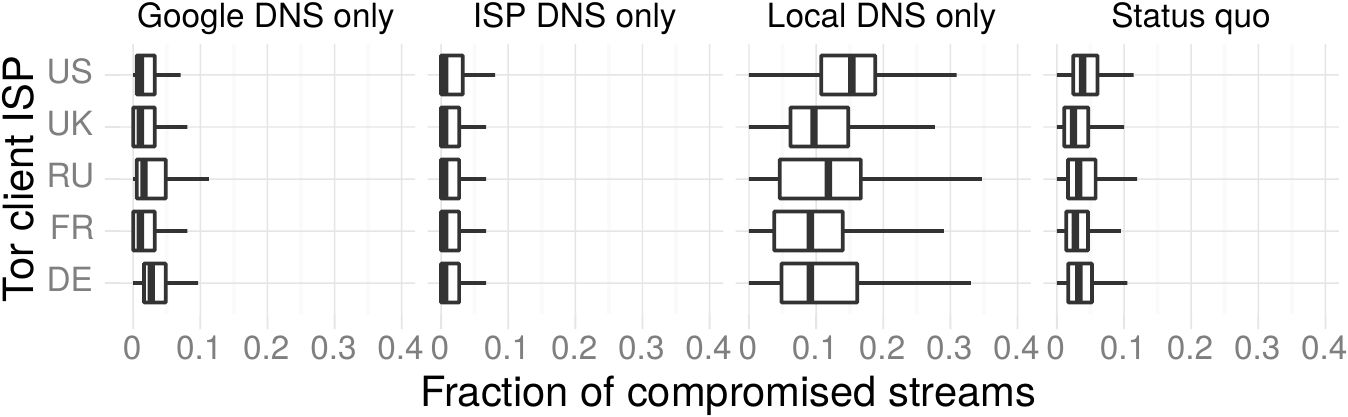}
	\label{fig:compromised-streams}
}
\subfigure[The time until simulated Tor clients got first compromised.]{
	\includegraphics[width=\linewidth]{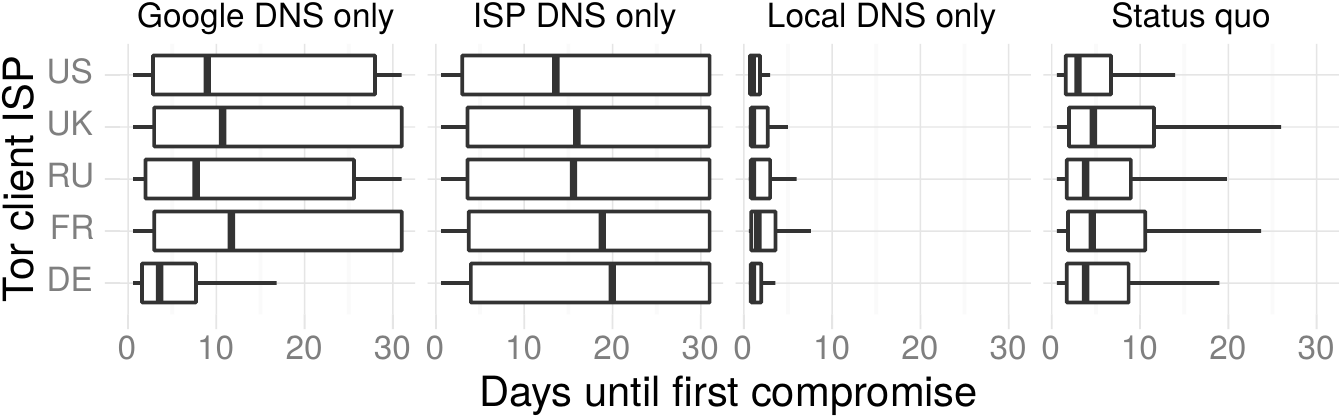}
	\label{fig:time-until-compromise}
}
\caption{The fraction of compromised streams and the time until first compromise
for our simulated Tor clients.  We placed these clients in five popular client
ASes in the U.S., the U.K., Russia, France, and Germany.  For exit relays, we
consider the status quo (on the very right) plus three hypothetical DNS
configurations for all exit relays.  We do not plot outliers beyond the box
plots' whiskers.  In both experiments, the safest configuration is ``ISP DNS
only,'' \ie have all exit relays use their ISP's DNS resolver.}
\end{figure}

Figures~\ref{fig:compromised-streams} and~\ref{fig:time-until-compromise}
illustrate our results as box plots.  Each figure contains four subfigures, one
for each DNS configuration.  Each box plot contains five rows, one for each Tor
client AS.  For clarity, we did not plot any outliers beyond the box whiskers.

Both figures show that the ``ISP DNS only'' setup is the safest for Tor users,
\ie, it exhibits on average the least number of compromised streams while also
on average counting the most days until compromise.  This setup is closely
followed by ``Google DNS only,'' the status quo, and finally ``Local DNS only,''
which fares worse than all other setups.  We expected ``ISP DNS only'' to do
best because if all exit relays use their ISP's resolvers, there is only one AS
to contend with on the egress side---the exit relay's.  The Google setup fares
similarly well most likely because of Google's heavily anycast infrastructure
which minimizes the number of AS hops.  The status quo does significantly better
than the ``Local DNS'' results, presumably because only around 12\% of Tor exit
relays actually do their own resolution.  The wide variance observed in
Figure~\ref{fig:time-until-compromise} for ``ISP DNS'' and ``Google DNS'' is due
to using 31 days as a placeholder for simulated clients who were never
compromised.  However, a safe configuration against AS-level adversaries, which
our figures capture, is not necessarily the best setup for Tor users.  For
example, ISP-provided DNS resolvers can be misconfigured, subject to censorship,
or simply be a forwarder to Google's resolver, which already serves numerous
exit relays and whose centralization poses a threat to the anonymity of Tor
users.  We will explore this trade-off in greater detail in
Section~\ref{sec:discussion}.

Interestingly, we find differences in our five client ASes.  These differences
are particularly striking in Figure~\ref{fig:time-until-compromise}.  
For example, for ``Google DNS only,'' the median time until compromise 
differs by around 7 days between DE and UK and around 8 days between 
DE and FR. For ``ISP DNS only,'' the median time until compromise differs by 
around 6 days between US and DE and around 5 days between US and FR.  
Also, we notice that DE fares worse than the others in the 
``Google DNS only'' scenario 
and better than the others in the ``ISP DNS only'' scenario.  
We conclude that the location of Tor clients matters and should be
considered in future traffic correlation studies.

\section{Discussion}
\label{sec:discussion}

In this section, we briefly discuss the ethics of our research and
possible ways to defend against \name attacks.

\subsection{Ethical considerations}
\label{sec:ethics}

Section~\ref{sec:load-freq} discusses how we set up an exit relay to
determine the number of DNS requests per five minute interval.  Since
our exit relay was forwarding traffic of real Tor users, we contacted Princeton
University's institutional review board (IRB) before running the
experiment.  Our IRB deemed that this research did not fall within the
realm of human subjects research.  In addition to contacting our IRB, we
adhered to The Tor Project's ethics guidelines~\cite{ethics-guidelines}.
Specifically, \first we ensured that we only collected data that is safe
to publish, \second we only collected data we needed, and \third we
limited the granularity of the data to minimize the likelihood of
reidentification.  The risk to Tor users of this experiment is negligible.
As for the benefits, by conducting this experiment, we can improve our
understanding of the risks that DNS poses to the anonymity of Tor users
and use this understanding to improve protection for Tor users in the
future.
Thus, we believe that the benefits of our experiment outweigh the risks.

\subsection{Defending against \name attacks}

We now discuss ways to defend against \name attacks.  We distinguish
between short-term solutions that can be implemented quickly
(Section~\ref{sec:short-term}), and long-term solutions that need significantly
more work (Section~\ref{sec:long-term}).

\subsubsection{Short-term solutions}
\label{sec:short-term}

Operators of exit relays face a dilemma: they must either operate their own
resolver, which exposes DNS queries to network adversaries; or, they must use a
third-party DNS resolver, which exposes DNS queries to a third party.  Clearly,
the goal is to minimize exposure of DNS requests, but there are several
dimensions to this.  In lieu of substantial DNS protocol improvements, we can
envision three extreme design points, in which \emph{all} exit relays use \first
Google's DNS resolver; \second their own, local resolver; or \third the resolver
provided by their ISP.  Table~\ref{tab:setup-comparison} summarizes the
important trade-offs for these three setups; the rest of this section discusses
these design points in more detail.  Table~\ref{tab:setup-comparison} shows
three attributes for our hypothetical DNS setups.  ``Network-level Protection''
is achieved if AS-level adversaries cannot easily monitor traffic, ``Avoiding
Centralization'' is achieved if there are many diverse resolvers instead of a
few central ones, and ``Response Quality'' is achieved if DNS responses arrive
quickly and correctly.

\begin{table}[t]
	\renewcommand{\tabcaptext}{A comparison between three design points for DNS
	resolver configuration, assuming all Tor exit relays use the setup in
	question.  Solid black circles are most desirable.}
	\topcap{\tabcaptext}
	\centering
	\begin{tabular}{l c c c}
	\toprule
	\textbf{Setup} &
	\begin{tabular}{@{}c@{}}\textbf{Network-level}\\\textbf{Protection}\end{tabular} &
	\begin{tabular}{@{}c@{}}\textbf{Avoiding}\\\textbf{Centralization}\end{tabular} &
	\begin{tabular}{@{}c@{}}\textbf{Response}\\\textbf{Quality}\end{tabular} \\
	\midrule
	All Google & \RIGHTcircle & \Circle & \CIRCLE \\
	All Local & \Circle & \CIRCLE & \CIRCLE\\
	All ISP & \CIRCLE & \RIGHTcircle & \RIGHTcircle \\
	\bottomrule
	\end{tabular}
	\bottomcap{\tabcaptext}
	\label{tab:setup-comparison}
\end{table}

If all exit relays were to use Google's public resolver, the company would
obtain metadata about the activity of all Tor users, which runs counter to Tor's
design goal of distributing trust.  We clearly should avoid this scenario.
Fifield \ea's~\cite{Fifield2015a} censorship circumvention system meek used to
use Google's cloud infrastructure to tunnel the traffic of censored users up
until May 2016~\cite{meek-shutdown}.  While the system was operational,
thousands of meek clients selected exit relays that use Google's public
resolver, which means that Google saw both traffic entering and, partially,
exiting the Tor network, allowing the company to mount \name attacks.

Next, consider a Tor network that only uses local resolvers.  In this case, Tor
is fully independent of third-party resolvers, at the cost of each iterative DNS
query being exposed to a diverse set of ASes in the network, allowing several
parties to learn the DNS queries of Tor users.

Finally, all exit relays could simply use their ISP-provided resolver.  This
would minimize the network exposure of DNS requests as resolvers are frequently
in the same AS as exit relays, and AS-level adversaries would be unable to
distinguish between DNS requests from exit relays and unrelated ISP customers.
This setup introduces the possibility of misconfigured and censored DNS
resolvers~\cite[\S~4.1]{Winter2014b}.  Besides, just a few ASes---OVH, for
example---host a disproportionate amount of exit relays, turning them into the
centralized data sinks that Tor aims to avoid.

Considering the above, we believe that exit relay operators should avoid public
resolvers such as Google and OpenDNS.  Instead, they should either use the
resolvers provided by their ISP, or run their own, particularly if the
operator's ISP already hosts many other exit relays.  Local resolvers can
further be optimized to minimize information leakage, by (for example) enabling
QNAME minimization~\cite{Bortzmeyer2016a}.

In addition to making recommendations to exit relay operators, we can remotely
influence the cache of each exit relay's resolver.  For example, using {\tt
exitmap}, we can continuously resolve potentially sensitive domains over each
exit relay, right before its TTL is about to expire.  In such a setup, an
attacker gains no advantage from observing DNS traffic from the exit relays
because the domain is always in every exit relay's resolver cache.  This
approach scales poorly, considering the potentially large number of domain names
that would need to be cached (recall that the long tail of unpopular sites are
most vulnerable to \name attacks), but it allows us to eliminate DNS-based
correlation attacks for a select number of sites.

Finally, Tor can fix the Tor clipping bug we discovered and consider
significantly increasing the minimum TTL for the DNS cache at exit relays to
make \name attacks less precise.  This adjustment requires finding the longest
acceptable TTL that does not have a notable negative detriment to user
experience.  Further, as soon as the clipping bug is fixed, website operators
of sensitive websites can opt to increase the TTL of their DNS records.

\subsubsection{Long-term solutions}
\label{sec:long-term}

Additional practical defenses are on the horizon.  Zhu \ea~\cite{Zhu2015a}
proposed T-DNS, which employs several TCP optimizations to transport the DNS
protocol over TLS and TCP.  The TLS layer provides confidentiality between exit
relays and their resolvers.  Finally, site operators whose users are
particularly concerned about safety should offer an onion service as an
alternative.  Facebook, for example, set up {\tt facebookcorewwwi.onion}.  When
connecting to the onion service, Tor users never leave the Tor network, and
hence do not need DNS (as long as the onion service does not embed non-onion
service content).

Deploying defenses against website fingerprinting attacks in Tor should be an
important long-term goal, as well.
Although growing the Tor network will help defend against \name attacks to some
degree, the most important change is to
deploy defenses against these attacks in the first place.  Since \name attacks
significantly increase precision of website fingerprinting attacks, defenses
should be designed to significantly reduce the recall of website fingerprinting
attacks, even when the website fingerprinting attack is configured to sacrifice
precision for recall.

\section{Conclusion}
\label{sec:conclusion}

In this paper, we have demonstrated how AS-level adversaries can use DNS
traffic from Tor exit relays to launch more effective correlation
attacks, which link the sender and receiver of traffic in the Tor
network.  We show how an attacker can use DNS query traffic to mount
perfectly precise \name attacks: Mapping DNS traffic to websites is highly
accurate even with simple techniques, and correlating the observed websites
with a website fingerprinting attack greatly improves the precision when
monitoring relatively unpopular websites. 

Given this more powerful fingerprinting method, we showed that the
threat of \name attacks against the Tor network is clear and present. We
first developed a method to identify the DNS resolver for each Tor exit relay,
and found that a set of exit relays comprising 40\% of all Tor exit
relay bandwidth use the Google public DNS servers. Although this
concentration of DNS query traffic reduces the expanse of ASes that can
see DNS query traffic emanating from exit nodes, this configuration
nonetheless gives a single administrative entity considerable visibility
into the traffic that is exiting the Tor network. Tor relay operators should
take steps to ensure that the network maintains more diversity into how
exit relays resolve DNS domains.

On the other hand, iterative DNS resolution from a Tor exit relay allows
a greater number of ASes to observe DNS queries from Tor exit relays.
For the Alexa most popular 1,000 websites, about 60\% of the DNS lookups
required to resolve the website's DNS domain name traversed ASes that
were {\em not} on the end-to-end network path between the client and the
resolved IP address for the webserver.  To mitigate the risk of
correlation attacks in light of this finding, we suggest that local DNS
resolvers on Tor exit relays implement privacy-preserving techniques
such as DNS QNAME minimization, which minimizes the amount of
information about the domain name that each iterative query contains.

Website fingerprinting attacks have long been a concern for the Tor
network. The attacks that we present in this paper show that, when
incorporating DNS query traffic, these attacks become even more accurate
and powerful. We hope these findings underscore the urgency of
eventually deploying strong defenses against fingerprinting attacks on
the Tor network.

We publish all our code, data, and replication instructions on our project page,
which is available online at \url{https://nymity.ch/tor-dns/}.

\section*{Acknowledgments}
We want to thank Jedidiah R. Crandall for providing infrastructure for our
measurements, Aaron Johnson for help with TorPS, Robayet Nasim for running
\name experiments, and Tom Ritter for providing
helpful feedback.
This research was supported in part by the Swedish Foundation for Strategic
Research grant SSF FFL09-0086; the Swedish Research Council grant VR 2009-3793;
the Swedish Internet Fund grant `Hoppet till Tor'; the National Science
Foundation Awards CNS-1540055 and CNS-1602399; and the Center for Information
Technology Policy at Princeton University.

\bibliographystyle{IEEEtranS}
\bibliography{references}
\balance

\end{document}